\documentclass[fleqn,usenatbib]{mnras}

\usepackage{newtxtext,newtxmath}

\usepackage[T1]{fontenc}

\DeclareRobustCommand{\VAN}[3]{#2}
\let\VANthebibliography\thebibliography
\def\thebibliography{\DeclareRobustCommand{\VAN}[3]{##3}\VANthebibliography}

\usepackage{graphicx}	
\usepackage{amsmath}	
\usepackage{pdflscape}
\usepackage[aboveskip=0pt]{caption}
\usepackage{float}
\usepackage{booktabs,siunitx}
\usepackage{caption}
\usepackage{subcaption}
\usepackage{multicol}

\title[Association of optical, UV and soft X-ray emission in AGNs ]{Association of Optical, Ultraviolet and Soft X-ray excess Emissions in AGNs}

\author[D. Nour and K. Sriram]{
D. Nour,$^{1,2}$\thanks{E-mail:nour.abb.dergham@gmail.com} and 
K. Sriram$^{1}$\thanks{E-mail:astrosriram@yahoo.co.in}
\\
$^{1}$Department of Astronomy, Osmania University, Hyderabad 500007, India\\
$^{2}$ Department of Physics, Faculty of Science, Al Baath University, Homs, Syria \\
}

\date{Accepted XXX. Received YYY; in original form ZZZ}

\pubyear{2015}

\begin{document}
\label{firstpage}
\pagerange{\pageref{firstpage}--\pageref{lastpage}}
\maketitle

\begin{abstract}
Soft X-ray emission (0.5--2.0 keV) plays a pivotal role in regulating the optical and UV emission in the AGNs. We collected a sample of 1811 AGNs from the SDSS database and obtained various parameters of Balmer lines, optical continuum, MgII line \& UV continuum and studied their dependencies on soft X-ray luminosity. Based on the linear regression analysis, we found that FWHM$_{\text{MgII}}$ $\propto$ FWHM$_{\text{H}\beta}^{0.554}$ suggesting that UV emission is arising from a region relatively outside the broad line region (BLR) associated to the H$\beta$ emission and found a strong correlation between optical and UV luminosities (L$_{\text{MgII}}$ $\propto$ L$_{\text{H}\beta}^{0.822}$). It was noticed that the dependency of optical continuum luminosities on soft excess changes with the redshift (L$_{\text{X}}$ $\propto$ L$^{0.596}_{5100\text{\AA}}$ for z < 0.5 and L$_{\text{X}}$ $\propto$ L$^{0.429}_{5100\text{\AA}}$ for z > 0.5). The FWHM components of H${\beta}$ and MgII core components were found to be virialized and is not affected by the soft excess emission whereas the wings of MgII display a dependency. We estimated a relation viz. L$_{\text{X}}$ $\propto$L$^{0.520}_{3000\text{\AA}}$ FWHM$^{0.525}_{\text{MgII}}$ and found to be well in agreement with a proposed physical scenario. All the derived relations were used to understand the inter-modulating association of the BLR and disc in the AGNs.  

\end{abstract}




\section{Introduction}
Accretion of matter onto the supermassive black holes (SMBHs) at the center of galaxies is the main source of power in active galactic nuclei (AGNs) causing their bolometric luminosity to be of order of 10$^{48}$ erg s$^{-1}$ \citep{woo2002active} which make them as one of the most energetic objects in the universe. The width of the optical emission lines are notable features in Type I AGNs and based on the broadness, they are classified into two sub-classes viz. narrow-line Seyfert galaxies (NLSy1) and broad-line Seyfert galaxies (BLSy1) \citep{zhou2006comprehensive,tarchi2011narrow} where the former is characterized by full width at half maximum (FWHM) of H$\beta$ $\leq$ 2200 km s$^{-1}$ and flux ratio of [OIII] to H$\beta$ to be $<$ 3 \citep{osterbrock1985spectra}. Broad emission lines are originating in the broad line region (BLR) \citep{osterbrock1989book,peterson1997introduction} that extends around tens of light-days from SMBH \citep{kaspi2005relationship,bentz2006reverberation} and are mainly due to photoionization by ultraviolet (UV) and X-ray emission originating in and above the accretion disc \citep{bahcall1969some,davidson1979emission}. However there are different suggestions about the structure of this region, some considered the emission to be due to accretion disc wind \citep{proga2000dynamics, kollatschny2003accretion,elvis2017quasar}, whereas \citep{netzer1993dust} considered the BLR to extend from the inner accretion disc to the inner torus and its emission originating within the sublimation radius. Other studies of the double-peaked broad emission lines suggested that BLR emission is originating from or close to the outer accretion disc \citep{strateva2003double,storchi2017double}.

The most frequently used broad emission lines to study the BLR at low redshift are H$\beta$ and H$\alpha$, but for redshifts higher than 1, MgII line becomes the line of interest as H$\beta$ and H$\alpha$ lines will shift out of the optical band. Although broad lines are supposed to originate from almost the same region but they have different properties and emission mechanisms, H$\beta$ and H$\alpha$ are recombination lines while MgII is dominated by collisional excitation \citep{guo2020understanding}. MgII emitting region is considered to be extended to larger radius compared to H$\beta$ and H$\alpha$ emitting regions \citep{salviander2007black,wang2009estimating,popovic2019structure,savic2020estimating}. \citet{popovic2019structure} stated that the region emitting MgII has more complex structure than the one emitting Balmer broad lines and in their model they considered the MgII broad line to originate from two sub-regions i.e. one contributes to the core line and the other contributes to the broad wings of the line.

The X-ray spectrum of an AGN is characterized by many physical and radiative components based on the energy bands, primarily the power-law component dominates the spectrum between 2 - 10 keV and originates from Comptonization of disc photons by a compact region smaller than the outer radius of the accretion disc known as corona \citep{haardt1991two} or a jet. Below 2 keV, many AGNs exhibit an excess of photons known as the soft excess component. The soft excess in the X-ray spectra (0.5-2.0 keV) is almost observed in a broad class of AGNs but many proposed mechanisms take the claim to explain the feature. Basically this feature is a residual emission when a power-law is extrapolated to higher X-ray continuum emission \citep{arnaud1985exosat}. The soft excess can be well fitted with a black body model with a corresponding temperature found to be around 0.1--0.2 keV across a broad range of AGN mass \citep[e.g.][]{gierlinski2004soft}. But this cannot be a thermal emission as it is too high for a standard accretion disc as proposed by \citet{shakura1973black}. Another strong interpretation of the soft excess is the Comptonization of the UV photons arising from the disc \citep{porquet2004xmm}. \citet{magdziarz1998spectral} suggested the existence of secondary corona between the disc and the hot corona that is cooler and has higher optical depth which in turns causes Comptonization of X-ray photons. \citet{tanaka2004partial} considered the soft excess to be due to the partial covering factor whereas \citet{ballantyne2001x} stated that reflection in the inner accretion disc might be a natural reason for this emission \citep[see also][]{crummy2006explanation}.

Studying X-ray emission of Type I AGNs provides us information about the processes occurring at the innermost regions of AGNs as we have a direct view of the central engine \citep{antonucci1993unified}. Many studies have been carried out to investigate the relation between optical emission lines of Type I AGNs and X-ray photon index. \citet{boller1995soft} and \citet{wang1996x} reported a significant correlation between the soft X-ray photon index ($\Gamma$s) and FWHM (H$\beta$) in NLSy1. \citet{ojha2020comparison} studied the relation between FWHM (H$\beta$) and hard X-ray photon index ($\Gamma$h) as well as the soft one ($\Gamma$s) for a sample of 139 NLSy1 and 97 BLSy1. They found a significant anti-correlation between ($\Gamma$s) and FWHM (H$\beta$) in BLS1 and observed a nominal correlation in case of NLS1, whereas in case of ($\Gamma$h) both sub-types show mild correlation. 

There are several attempts to study the evolution of X-ray spectra of AGNs over time. \citet{brinkmann1997broad} and \citet{yuan1998broad} found a strong correlation between X-ray spectral index and redshift for a large sample of radio loud and radio quiet where z $<$ 2. Similar relation was found by \citet{kelly2007evolution} for redshift range of 0.1--3 and for high redshift AGNs 3.7 $>$ z $>$ 6.28 \citep{bechtold2003chandra}. On the other hand \citet{page2003serendipitous}, \citet{young2009fifth}, and \citet{nanni2017x} did not reported any evidence of evolution. \citet{kelly2007evolution} stated that not all correlations between photon index and redshift indicates spectral evolution with time, but it might be due to different magnitude of contribution of soft excess and Compton hump components at different redshifts. However the evolution of the relation between soft X-ray and UV luminosities has been investigated by \citet{risaliti2015hubble}, they concluded that this relation is not changing over time and therefor can be used to estimate the cosmological constants.

In this study we explored the association between BLR emission lines (H$\beta$ \& MgII) parameters and soft X-ray luminosity in the energy band (0.5--2.0 keV) (L$_{\text{X}}$). We would like to explore how the soft excess or the reflection component originating in the keplerian disc close to the SMBH affects the radiative and dynamic configuration of the BLR. \citet{liu2016x} modeled the X-ray spectra of AGNs for one of the largest samples, with a consistent physical model which accounts for the soft excess. As not many works studied the relation between the soft X-ray luminosity in the energy band (0.5-2keV) and optical/UV properties for a large sample, we cross-matched the SDSS dr 14 catalogue \citep{paris2018sloan} with \citet{liu2016x} to finally get the large sample presented in the work. The present study is the one of the largest collected samples to constrain the relations between optical / UV continuum parameters \& L$_{\text{X}}$ and investigate the strength of these correlations. We also study  the evolution of correlation strengths over cosmic time (i.e. z).

\section{DATA SELECTION AND REDUCTION}
The quasars catalogue \citep{paris2018sloan} includes all SDSS-IV/eBOSS objects that are confirmed to be quasars by having luminosities M$_{\text{i}}$ [z = 2] $<$ -20.5 and show at least one emission line with FWHM $>$ 500 km s$^{-1}$ or have an absorption feature. It contains 526 356 quasars for which we obtained X-ray counterparts by cross-matching \citet{paris2018sloan} catalogue with X-ray catalogue given by \citet{liu2016x} within a radius of 5 arcseconds. This catalogue contains 8445 point-like X-ray sources detected in the XMM-XXL north survey \citep{pierre2016xxl}, 2512 of them have reliable redshift measurements and the total coverage area of this survey is about 50 deg$^{2}$ divided into two fields pertaining equal size. By limiting our choice to Type I AGNs, this resulted in a sample of 1819 sources and we excluded 8 AGNs with z $>$ 3 due to high dispersion, hence the final sample studied in this paper includes 1811 AGNs.
Due to the wavelength coverage of SDSS spectra (3610--10140 \AA), not all spectra cover both MgII and H$\beta$ lines (i.e. 
115 objects of the sample show H$\beta$ line without MgII line in their spectra, whereas 1267 spectra show only MgII and the rest 429 objects show both H$\beta$ and MgII). However, the number of sources used in each sub-sample studied in the present work are listed in Table. 1
\begin{table}
\caption{Number of sources of each sub-sample studied in this paper}
\label{tab:T1}
\setlength\tabcolsep{3 pt}
\setlength{\cmidrulekern}{2em}
\centering
\begin{tabular*}{\columnwidth}{%
    l c }
\toprule

Sub-sample& No. of sources\\
\midrule
1) H$\beta$ vs MgII (FWHM \& luminosity) & 429 \\
2) L$_{\text{X}}$ vs L$_{5100\text{\AA}}$ & 468 \\
3) L$_{\text{X}}$ vs H$\beta$ (FWHM \& luminosity) & 543 \\
4) L$_{\text{X}}$ vs L$_{3000\text{\AA}}$ & 1515 \\
5) L$_{\text{X}}$ vs MgII (FWHM \& luminosity) & 1695 \\
6) L$_{\text{X}}$ vs L$_{3000\text{\AA}}$ vs FWHM (MgII) & 1492 \\

\hline
\end{tabular*}
\end{table}

To obtain the optical / UV properties of the sample, we downloaded the processed spectra from Sloan Digital Sky Survey (SDSS) DR14 database and later we analyzed each spectrum individually by applying the following corrections 
\citep[for more details see][]{sriram2022influence}.

1. We first corrected each spectrum for galactic extinction using \citet{fitzpatrick1999correcting}  parameters and the galactic extinction maps \citet{schlegel1998maps}.

2. Second step was to correct wavelength and flux for redshift where
$\lambda_ {\text{rest}}$ = $\lambda_ {\text{observed}}$ / (1+z) , f$_{\text{rest}}$ = f$_{\text{observed}}$ * (1+z).

3. Host galaxy contribution subtraction: It is necessary to eliminate the host galaxy contribution in the optical spectra in order to perform an accurate study for the central engine. \citet{berk2006spectral} stated that an AGN spectrum can be reconstructed as linear combination of eigenspectra. Therefor by performing principal component analysis (PCA) technique, we divided each AGN spectrum into two parts, host galaxy part (GAL) and central engine part (QSO) using quasars and galaxy eigenspectra given by \citet{yip2004distributions,yip2004spectral} as given in the relation
\begin{equation}
f(\lambda)=\sum_{i=1}^{m} a_i Q_i(\lambda) + \sum_{j=1}^{n} b_j G_j(\lambda)
\end{equation}

Where f($\lambda$) is the best-fitting flux, a$_{\text{i}}$ and b$_{\text{j}}$ are eigencoefficients and  $Q_\text{i}$ and $G_\text{j}$ are QSO and galaxy eigenspectra respectively. As the properties of an AGN spectrum can be recovered by the first few eigenspectra, we chose the number of QSO eigenspectra m=10 and GAL eigenspectra n=5.

4. The continuum below the emission lines was fitted separately for different regions (i.e. 4000--5500 \AA \, for H$\beta$ and [OIII] emission lines, 6300--6800 \AA \,for H$\alpha$, [NII], and [SII] emission lines, and 2650--3050 \AA \, for MgII). For each of these regions, we considered the continuum windows given in \citet{kuraszkiewicz2002emission} then interpolate these points by fitting a power law that determines the continuum level in order to subtract it.
\begin{figure}
\centering
\includegraphics[width=\linewidth]{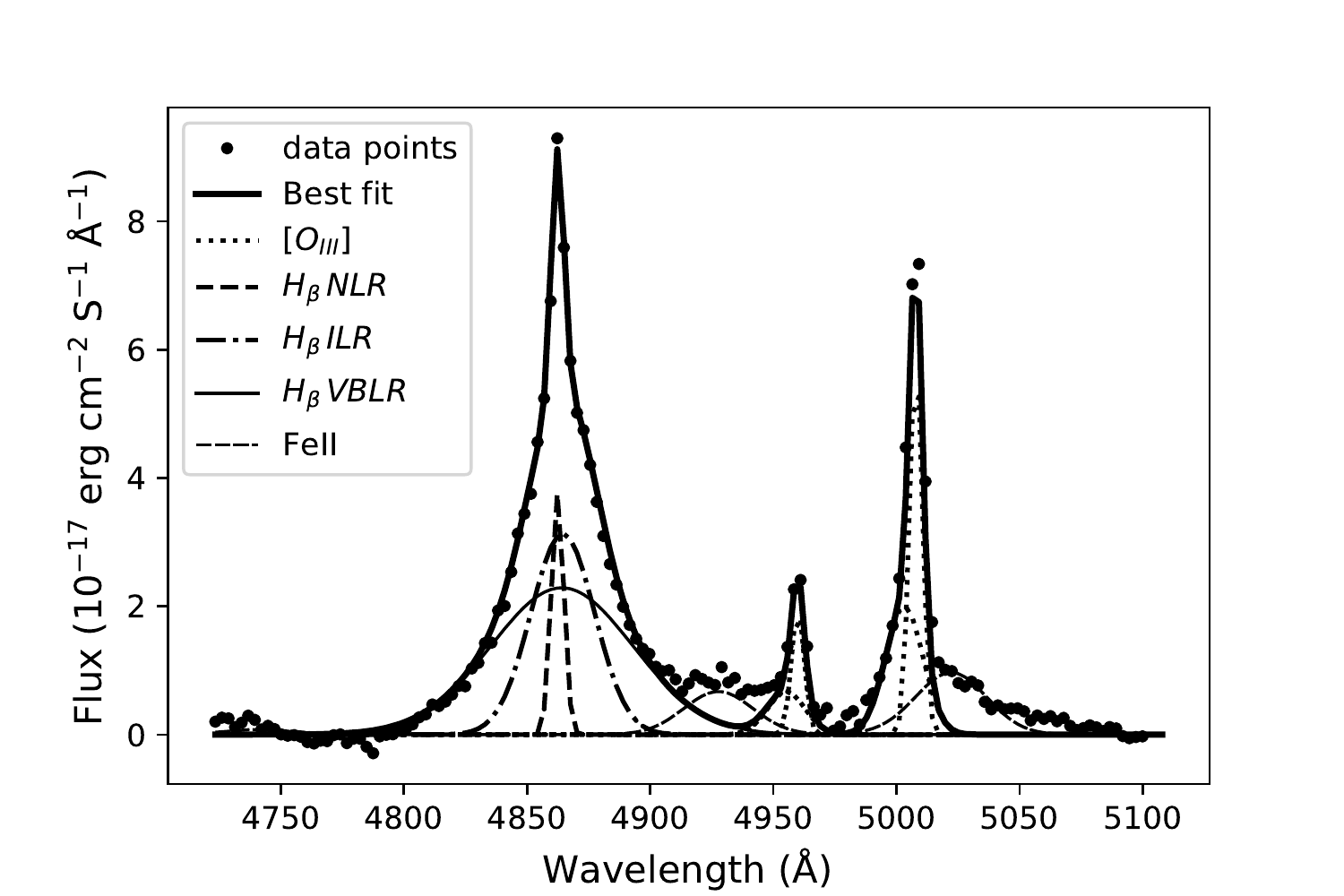}
\caption{\label{fig:1}Example of the best of emission lines near H$\beta$ for a source SDSS J020853.28-043353.5. Different components are shown in different line markers (see the legend).}
\end{figure}
\begin{figure}{}
\centering
\includegraphics[width=\linewidth]{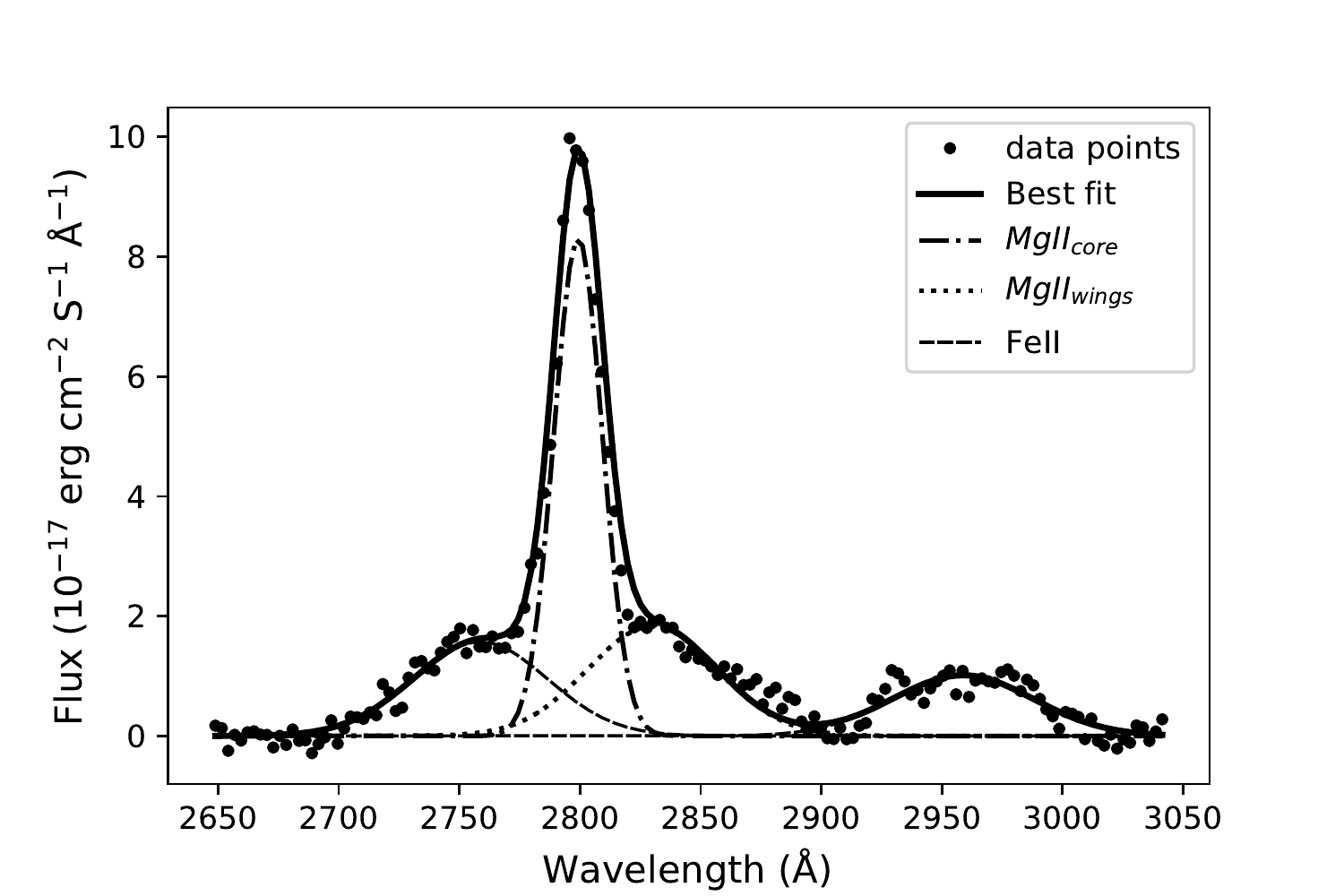}
\caption{\label{fig:2}Best fit of a MgII emission line ina source SDSS J022452.20-040519.5. Different components are shown with different markers (see the legend)}
\end{figure}

5. Emission lines fitting: Before fitting the emission lines we have to overcome the optical FeII feature that has a strong contribution in the region of 4000–5500 \AA\, and hence affects the H$\beta$ and OIII line estimation.
Using an online software given by \citet{kovavcevic2010analysis}  and \citet{shapovalova2012spectral}, we fitted the optical FeII and subtracted it from the spectrum. To fit the broad emission lines we considered a two-component model \citep{popovic2004contribution,bon2009contribution} that suggested the BLR emission to be modelled by two components originating from a two kinematically different regions i.e. intermediate line region (ILR) that contributes to the core of the emission line and very broad line region (VBLR) that form the wings of the line and is closer to the SMBH.
H$\beta$ line was fitted using three Gaussian components viz. one represents emission for the narrow line region (NLR), and remaining two for the BLR emission (ILR and VBLR). 
MgII line overlaps with the UV FeII which was fitted and subtracted using the method described by \citet{kovavcevic2015connections} and later MgII line was fitted using two Gaussian components viz. one for the core and another to fit the wings. Figs. 1 and 2 show the best-fits of H$\beta$ and MgII lines and the respective insets display the various components. 

6. The last step was to calculate all required parameters of emission lines components such as fluxes, luminosities, equivalent widths (EWs), FWHMs as well as the continuum luminosity at 5100 \AA\, and 3000 \AA.

\citet{liu2016x} analyzed the X-ray spectra in their catalogue using the Bayesian X-ray Analysis software \citep{buchner2014x} in order to fit the spectra with a model consisting of three main components (a) BNTORUS model \citep{brightman2011xmm} to fit the power-law continuum, absorption and Compton scattering features; (b) the PEXMON model \citep{nandra2007xmm} which represents the reflection component; (c) a soft scattering component to account for the soft X-ray excess in AGNs. For our study we adopted the soft X-ray fluxes reported in \citet{liu2016x} to calculate the soft X-ray luminosities for a sample of 1811 AGNs in order to constrain the association of the reflection component emitting from the disc on the H$\beta$ and MgII emitting regions. These soft X-ray luminosities are corrected for galactic absorption and no other components, such as warm absorbers, have been taken into account.
The cosmological parameters used for our calculations are $\Omega_M$ = 0.3, $\Omega_\Lambda$ = 0.7, and $\Omega_k$ = 0, H$_0$ = 70 km s$^{-1}$ Mpc$^ {-1}$.
The derived optical and UV parameters used in this paper are listed in Table 2 along with the estimated X-ray luminosity. In Table 3, we listed Spearman's correlation coefficient $\rho$ and probability P between the different parameters.
\section{Results and Discussion}

We performed the correlation studies in order to understand the association of optical/UV emitting regions and soft X-ray component which origin is still debated. We also studied the correlations between soft X-ray luminosity (L$_\text{X}$) and continuum luminosities at 5100\AA\, and 3000\AA\, as well as FWHMs of broad components of H${\beta}$ and MgII. We determined the regression lines and Spearman's correlation coefficients ($\rho$) for all correlations.

\subsection{Optical and UV properties of the sample}
The relation between FWHMs of broad components of Balmer lines (H$\alpha$ and H$\beta$) have been studied for different samples across a broad range of redshifts. \citet{greene2005estimating} studied a sample of 299 AGNs with z < 0.3, \citet{mejia2016active} chose a sample of 39 AGNs with a redshift up to z = 1.6 and both studies found a strong correlation between Balmer lines FWHMs consistent following a one is to one relation. They suggested that both of these lines are emitted from the same region.
However there exist different assumptions in case of the relation between FWHMs of MgII line and Balmer lines. Some studies indicate that the MgII and H$\beta$ have the same width.  \citep{mclure2002measuring,onken2008improved,shen2008biases}.   \citet{corbett2003emission} observed a different result claiming that MgII is broader than H$\beta$. However other studies showed that Mg\text{II} line is narrower than H$\beta$ line \citep{salviander2007black,wang2009estimating,vietri2020super} and this difference in width was considered as an indication that MgII is emitted from a region located at a larger radius compared to H$\alpha$ and H$\beta$ emitting region. We investigated the correlation between FWHMs of H$\beta$ and MgII broad components for 429 sources that have both the lines in their spectra. The fitted regression line resulted in a significant correlation ($\rho$ = 0.58) between FWHM of H$\beta$ and Mg\text{II} with a probability P = 4.3e--32  (see Fig. 3). The best fit by a linear relation is shown in equation 2 with a slope = 0.55 indicates that the width of Mg\text{II} emission line is narrower than the H$\beta$. A similar slope was also obtained by \citet{shen2012comparing} (slope = 0.57) for a sample of 60 intermediate redshift quasars. Hence our sample supports the previous studies that the MgII emitting region should be located at a larger radius than the region, emitting Balmer lines.
\begin{equation}
\log(\text{FWHM}_{\text{MgII}})=(0.554\pm0.033) \log(\text{FWHM}_ {\text{H}\beta)})+(1.561\pm0.118)
\end{equation}
\begin{figure}
\centering
\includegraphics[width=\linewidth]{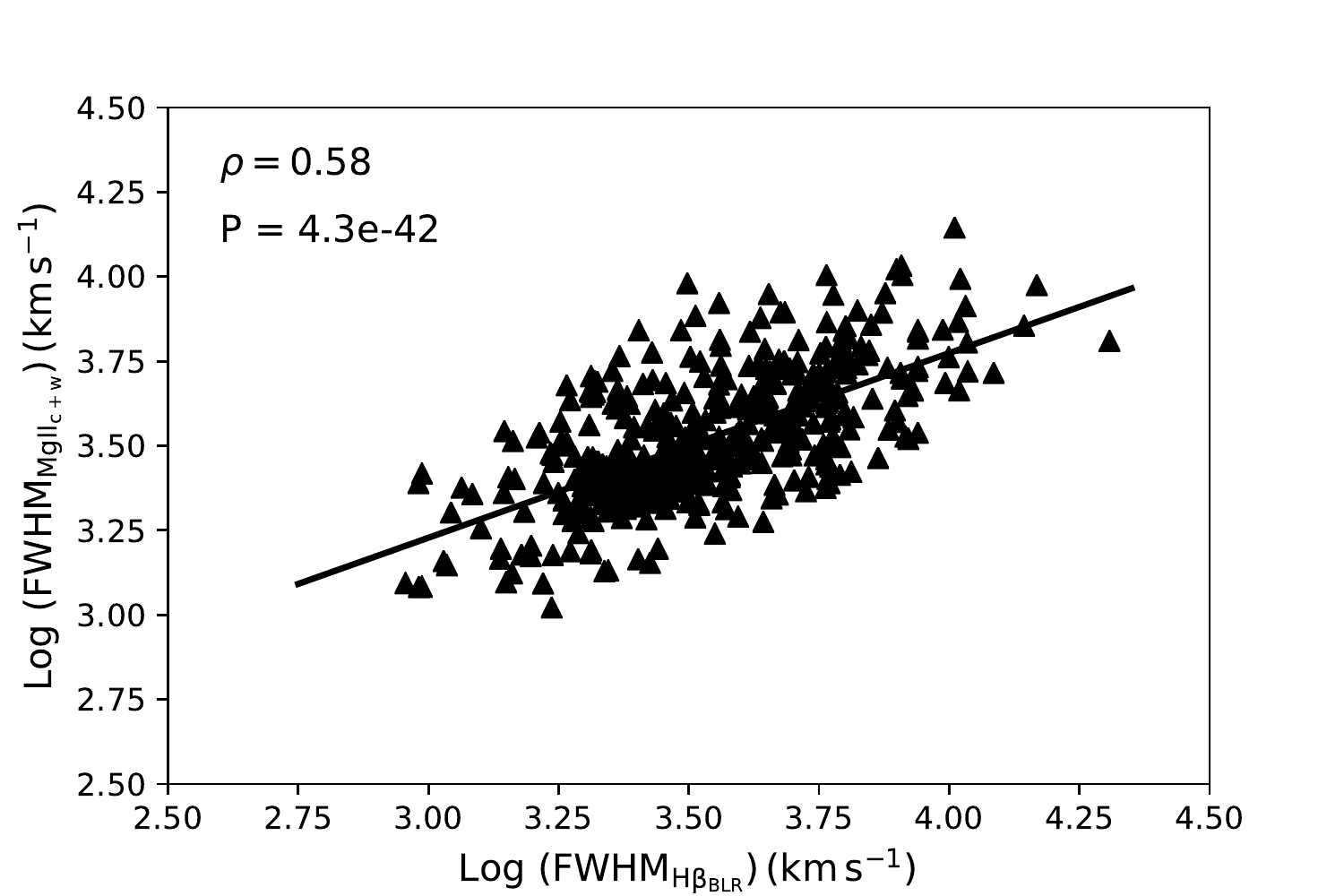}
\caption{\label{fig:3}The relation between FWHM$_{\text{MgII}}$ and FWHM$_{\text{H}\beta(\text{BLR})}$. The solid line shows the linear best-fit.}
\end{figure}
We also studied the correlation between the luminosities of the two emission lines (Fig. 4), and found them to correlate strongly $\rho$ = 0.83 with a slope = 0.822$\pm$0.028. This strong correlation indicates that both optical and UV emissions of the BLR are affected similarly by the ambient processes occurring near the central engine.

\begin{equation}
\log(\text{L}_{\text{MgII}})=(0.822\pm0.028) \log(\text{L}_{\text{H}\beta})+(7.742\pm1.192)
\end{equation}

\begin{figure}
\centering
\includegraphics[width=\linewidth]{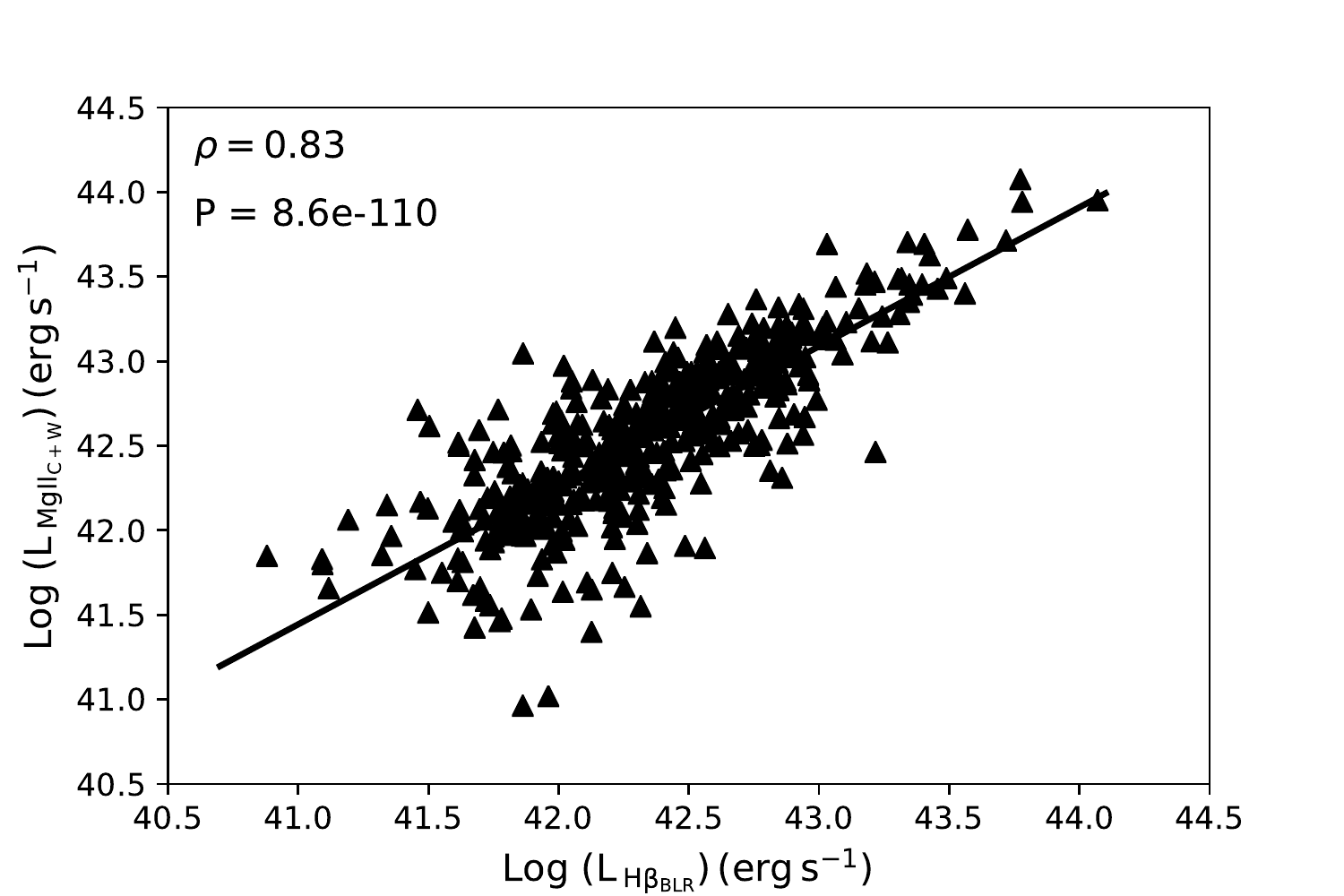}
\caption{\label{fig:4}The relation between L$_{\text{MgII}}$ and L$_{\text{H}\beta(\text{BLR})}$. The solid line is the best-fit for the data.}
\end{figure}

\subsection{Relation between Soft excess and Optical continuum emission components}
\subsubsection{Soft X-ray luminosity vs Optical continuum luminosity}
The optical and UV emissions in AGNs arise from both the BLR and the accretion disc \citep{padovani2017active} and possibly the accretion flow is the source of optical and UV continuum emission \citep{koratkar1999ultraviolet}. We studied the correlation between L$_{\text{X}}$ and optical continuum luminosity at 5100\AA\, (L$_{5100\text{\AA}}$). We found L$_{\text{X}}$ to correlate with L$_{5100\text{\AA}}$ ($\rho$= 0.63) (Fig. 5), with a probability $\ll$ 0.00001. This indicates that the soft excess feature possibly produced by the reflection of hard X-rays from the accretion disc, causes an increase in the the optical continuum emission.
The best-fitting line resulted in the following equation for AGNs z < 1.
\begin{equation}
\log(\text{L}_{\text{X}})=(0.596\pm0.035) \log(\text{L}_{5100\text{\AA}})+(17.094\pm1.563)
\end{equation}
\begin{figure}{}
\centering
\includegraphics[width=\linewidth]{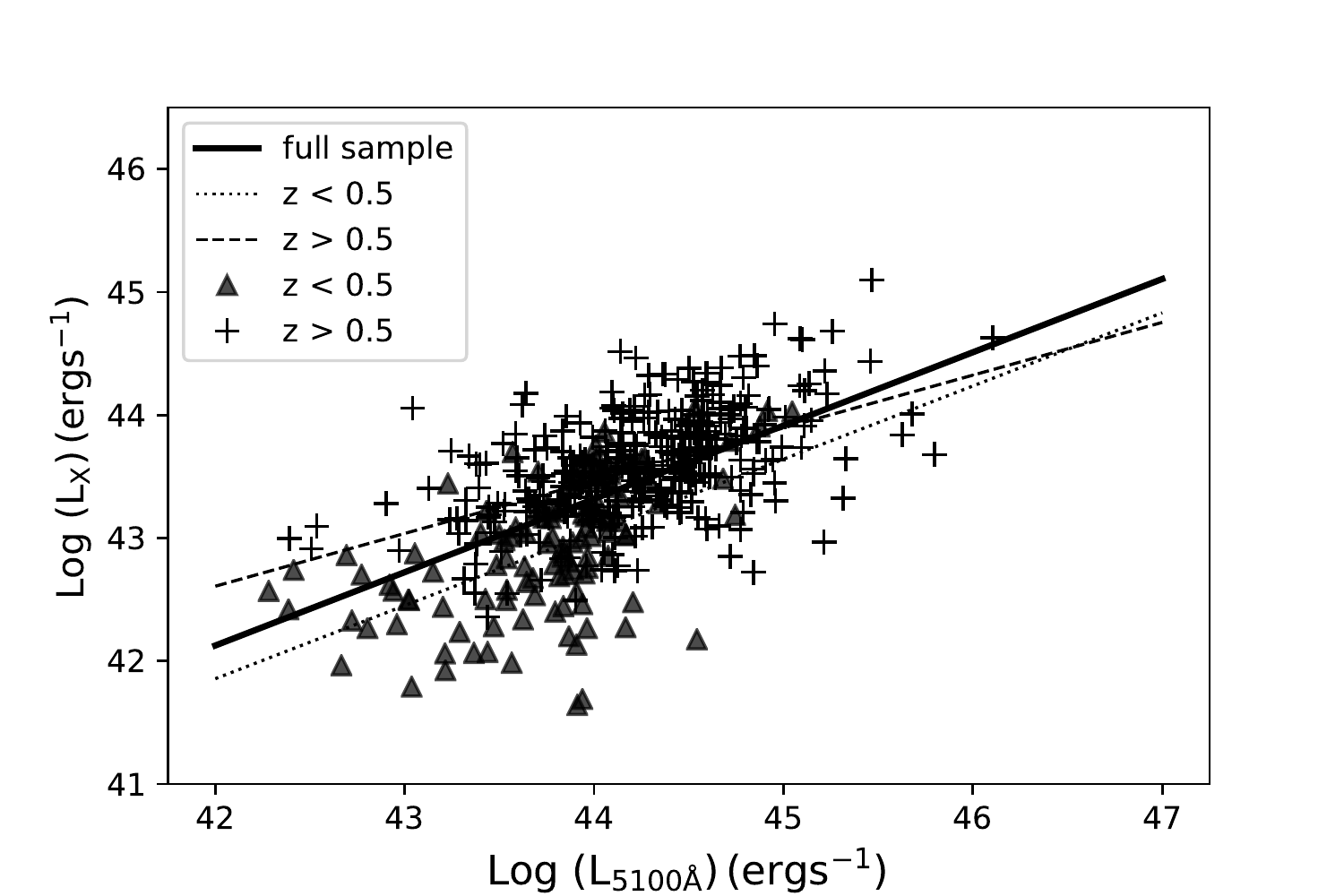}
\caption{\label{fig:5}Soft X-ray luminosity versus L$_{5100\text{\AA}}$. The solid line shows linear best-fit for the full sample, dotted line and dashed line display the linear best-fits for sources z < 0.5 (represented by triangles) and z > 0.5 (plus markers), respectively.}
\end{figure}
\begin{figure}{}
\centering
\includegraphics[width=\linewidth]{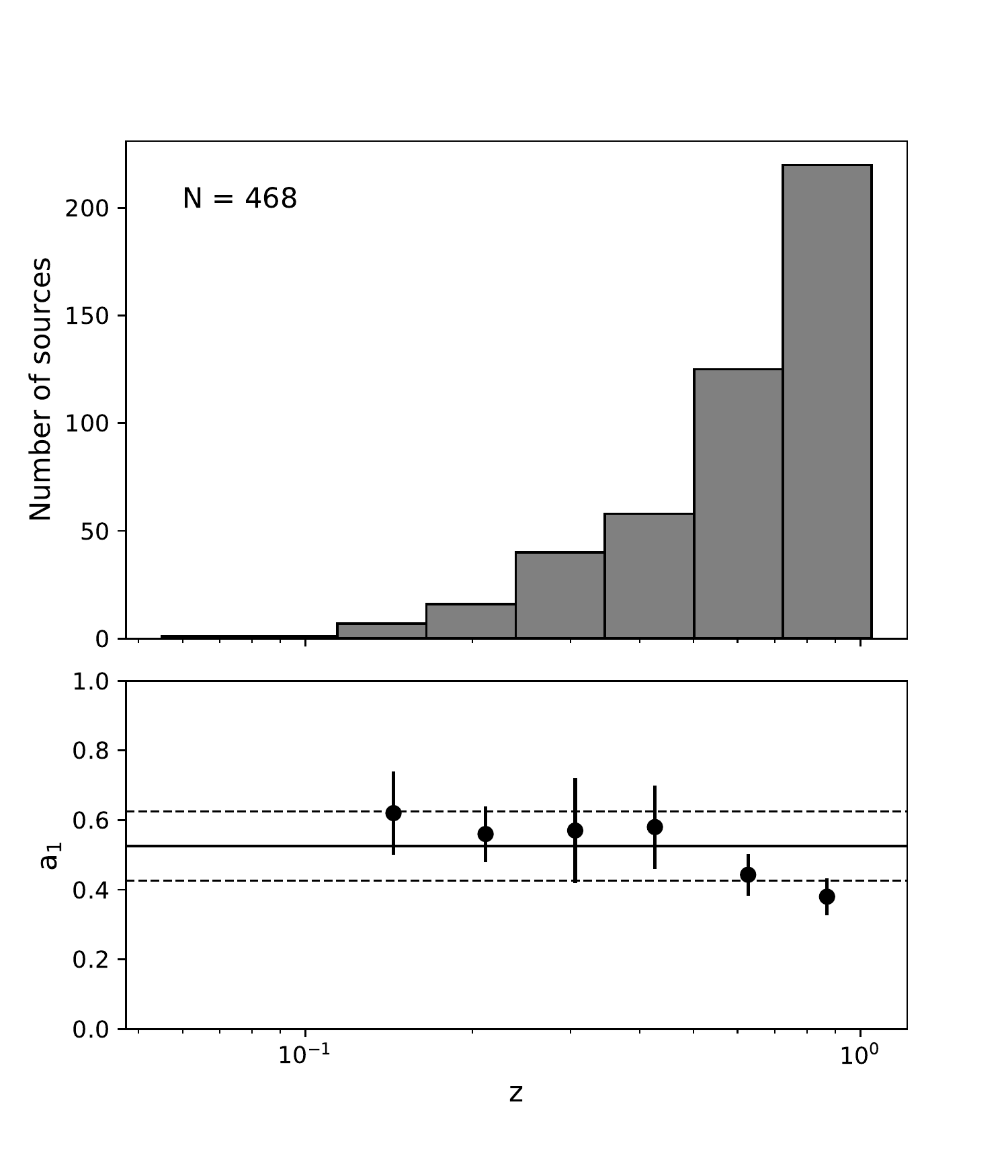}
\caption{\label{fig:5} Upper panel: redshift distribution of the sample of 468 AGNs for which both L$_X$ and L$_{5100\text{\AA}}$ were calculated. Lower panel: best-fitting values of the slopes in each redshift bin. The horizontal solid line represents the average of slopes obtained in each redshift bin $\overline{a_1}$= 0.525$\pm$0.081, the two dashed lines indicate the $\pm$1$\sigma$ region.}
\end{figure}
To check whether equation (4) displays any change with respect to z, we divided the sample into two parts $<$ z = 0.5 and z $>$ 0.5. \\
For  z $<$ 0.5, $\rho$ = 0.58
\begin{equation}
\log(\text{L}_{\text{X}})=(0.575\pm0.080) \log(\text{L}_{5100\text{\AA}})+(16.866\pm3.55)
\end{equation}
For  z $>$ 0.5, $\rho$ = 0.52
\begin{equation}
\log(\text{L}_{\text{X}})=(0.429\pm0.037) \log(\text{L}_{5100\text{\AA}})+(24.540\pm1.669)
\end{equation}
As shown in equations (5) and (6) (see Fig. 5), the strength of correlation was almost the same below and above z = 0.5. We noticed a small variation in the observed slopes as depicted from the best-fits. BLR of AGNs with z $>$ 0.5 are relatively least affected by the soft excess when compared to the BLR of AGNs with z $<$ 0.5. To confirm our results, we divided the sample into redshift bins with $\Delta$ log z = 0.16 (Fig. 6-upper panel). Then we fitted the data points in each bin with equation of the form L$_{\text{X}}$= a$_1$ log L$_{5100\text{\AA}}$ + b$_1$. We ignored the first two bins as the number of sources is $<$ 4. In lower panel of Fig.6, we display the relation between slopes of each bin and the redshift. We noted a slight decrease in slopes values for sources at z $>$ 0.5. A larger sample across the redshift would be helpful to reinforce the observed variation.   

However evolution studies using H$\beta$ can not give robust results as it is limited to sources z $<$ 1, because for sources with z $>$ 1, H$\beta$ will be shifted out of the optical spectrum range.

\subsubsection{L$_{X}$ vs L$_{\text{H}\beta}$ \& FWHM$_{\text{H}\beta}$ }
The broad component of H$\beta$ is considered to arises from the BLR that is considered to be relatively closer to the  central engine as per the unified model of AGNs. Previous studies confirmed the existence of a strong correlation between hard X-ray luminosity in 2--10 keV and the luminosity of the broad component of H$\beta$ \citep{bianchi2009caixa,piconcelli2005xmm,jin2012combined,sriram2022influence}. Assuming that the soft X-ray luminosity is due to the reflection mechanism over the disc, the hard X-ray and soft X-ray luminosities too are correlated. Hence we also found a strong correlation ($\rho$= 0.63) between the luminosity of BLR component and the soft excess (0.5--2.0 keV),  the best-fit line is shown in equation 7. This strong correlation continues to exist for other components of the broad emission line (ILR \& VBLR; see Fig. 7). The tight correlation indicates that the reflection component (i.e. soft excess) clearly influences the radiative and geometrical configuration of the BLR. We investigated the evolution of L$_{\text{X}}$-L$_{\text{H}\beta}$ relation  with redshift for ILR, VBLR and BLR by applying linear fit of the form log L$_{\text{X}}$= a$_2$ log L$_{\text{H}\beta}$ + b$_2$ for sources in redshift bins of 0.16 and did not find any significant change in the slopes values as shown in lower panels of Fig. 7. Since the variation is independent of z, this relation can be useful for cosmological studies but a larger sample is mandatory to confirm this result.


\begin{equation}
\log(\text{L}_{\text{X}})=(0.636\pm0.028) \log(\text{L}_{\text{H}\beta \text{BLR} })+(16.558\pm1.185)
\end{equation}
\begin{figure*}{}
\centering
\includegraphics[width=\linewidth]{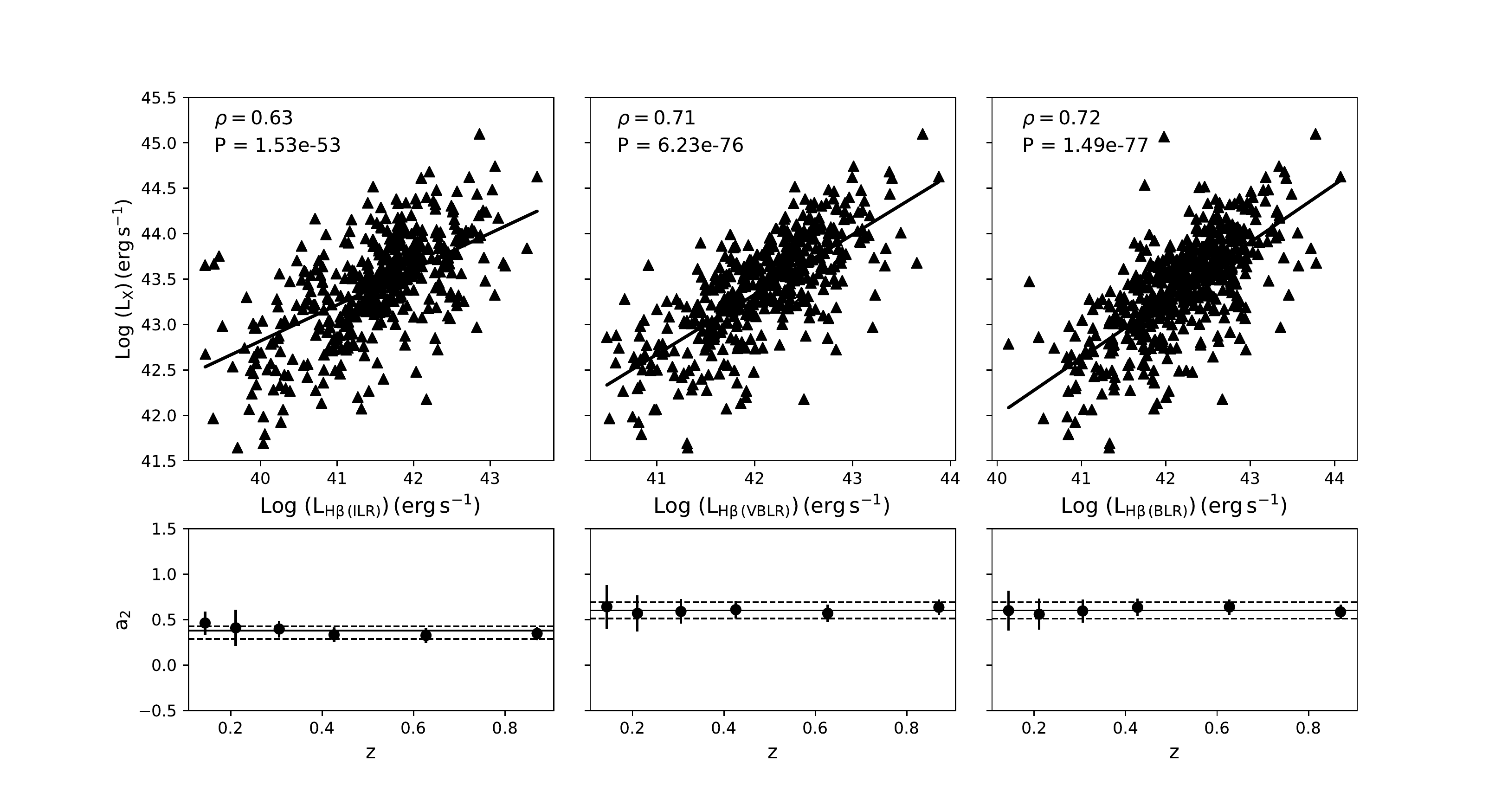}
\caption{\label{fig:6}Upper panels : The correlation between soft X-ray luminosity and luminosity of H$\beta$ components (i.e. ILR, VBLR, and BLR from left to right panels) along with linear best fits shown with lines. Lower panels show the best-fitting values of the slopes in each redshift bin. The horizontal solid lines show the average values of slopes obtained in each redshift bin (0.381$\pm$0.083; 0.610$\pm$0.071; 0.602$\pm$0.072), the dashed lines in each panel indicate the $\pm$1$\sigma$ region.}

\end{figure*}
\begin{figure*}{}
\centering
\includegraphics[width=\linewidth]{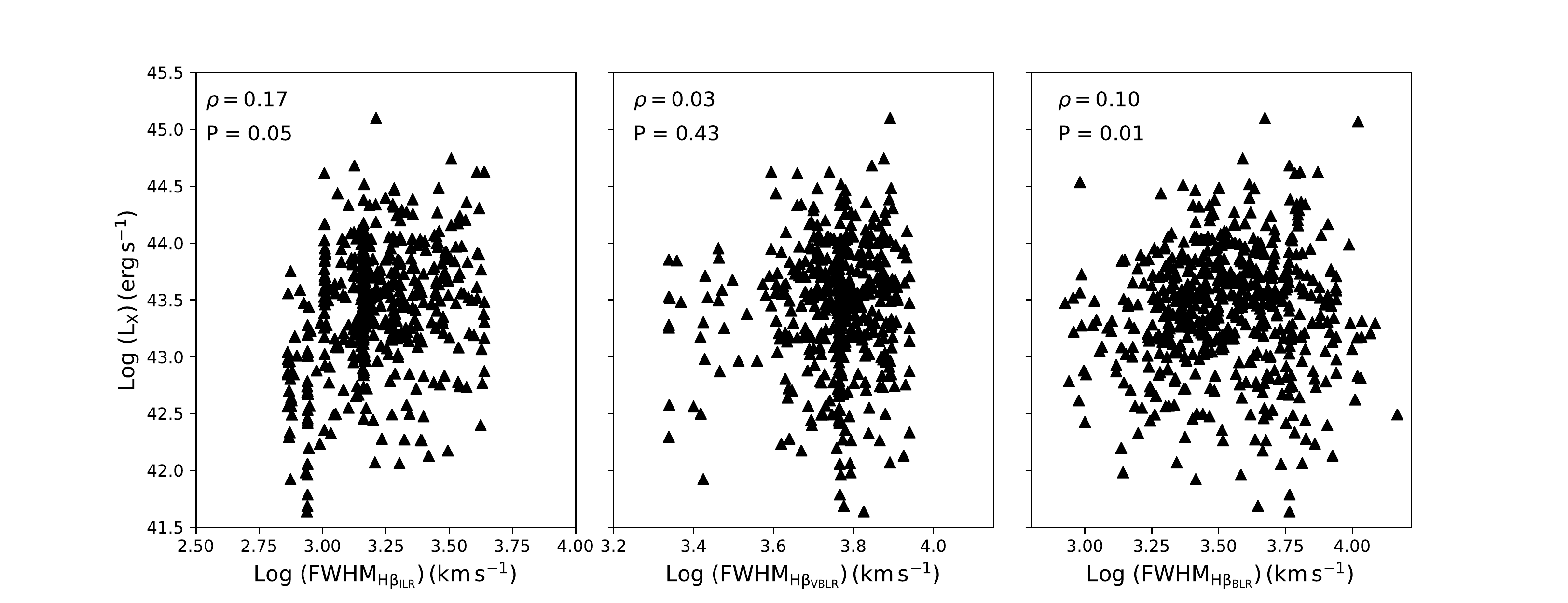}
\caption{\label{fig:7}Soft X-ray luminosity versus FWHMs of H$\beta$ components (i.e. ILR, VBLR, and BLR from left to right panels).}

\end{figure*}
The FWHM of H$\beta_{\text{BLR}}$ did not exhibit any correlation with L$_{\text{X}}$ and this agrees with the result reported by \citet{grupe2010simultaneous} for a sample of 92 bright soft X-ray sources. \citet{bianchi2009caixa} with a sample of 103 AGNs found a similar result, thus they considered the strong correlation between FWHM of H$\beta$ and concluded that the ratio between soft to hard X-ray luminosity is solely due to decrease in hard X-ray luminosity rather than increasing in the soft X-ray luminosity. 
FWHM of an emission line is affected by the gravity of the SMBH existing at the center of an AGN, this effect causes faster rotation of the surrounding material that have smaller radius compared with ones at larger radius (v$\approx(GM/R)^{1/2}$). As H$\beta$ emission line is broader than other Balmer lines as well as MgII emission line, therefor it is originating from a portion of the BLR closer to the SMBH than other broad emission lines and it shows virialization in the whole profile as expected from a ideal Keplerian motion \citep{popovic2019structure}. Hence its broadening has to be primarily due to the gravitational effect of the SMBH. We tested this suggestion for our sample by plotting the ILR component that fits the core of H$\beta$ line and the VBLR component which used to fit the wings versus L$_{\text{X}}$ (Fig. 8). We did not find any significant correlation, therefore we conclude that the whole line profile of H$\beta$ is virialized. The non-existence of a correlation between FWHMs of different Balmer components and soft X-ray luminosity indicates that this reflected emission does not affect the kinematics of the portion of BLR that emits H$\beta$ line. 

\subsection{Relation between soft excess and UV emission}
\subsubsection{Soft X-ray luminosity vs UV continuum luminosity}
The relation between L$_{\text{X}(\text{2keV})}$ and UV luminosity at 2500 \AA\, has been studied for sources up to z $\approx$ 6 over different magnitudes of luminosity. The largest studied samples include a sample of 545 sources \citep{lusso2010x} observed by XMM-Newton survey and a sample of 333 quasars \citep{steffen2006x} observed by different X-ray and optical surveys. \citet{young2009x} also studied a sample of 350 sources using the data from SDSS and XMM-Newton surveys and \citet{risaliti2015hubble} combined the data from previous studies to form the largest sample with 808 sources. All these studies confirm a non-linear relation between L$_{\text{X}(\text{2keV})}$ and L$_{2500\text{\AA}}$ with a slope ranging between 0.5--0.7, suggesting the existence of a universal mechanism connecting the coronal emission to accretion disc emission. \\
For the sample analyzed in this paper, we have obtained the UV luminosity at 3000\AA\ (L$_{3000\text{\AA}}$) along with L$_{\text{X}}$ for a sample of 1515 AGNs. We investigated the relation between these two luminosities and found a strong correlation between them ($\rho$ = 0.7; Fig. 9). The best-fit regression line resulted in the following relation.
\begin{equation}
\log(\text{L}_{\text{X}})=(0.557\pm0.014) \log(\text{L}_{3000\text{\AA}})+(18.943\pm0.655)
\end{equation}
\begin{figure}
\centering
\includegraphics[width=\linewidth]{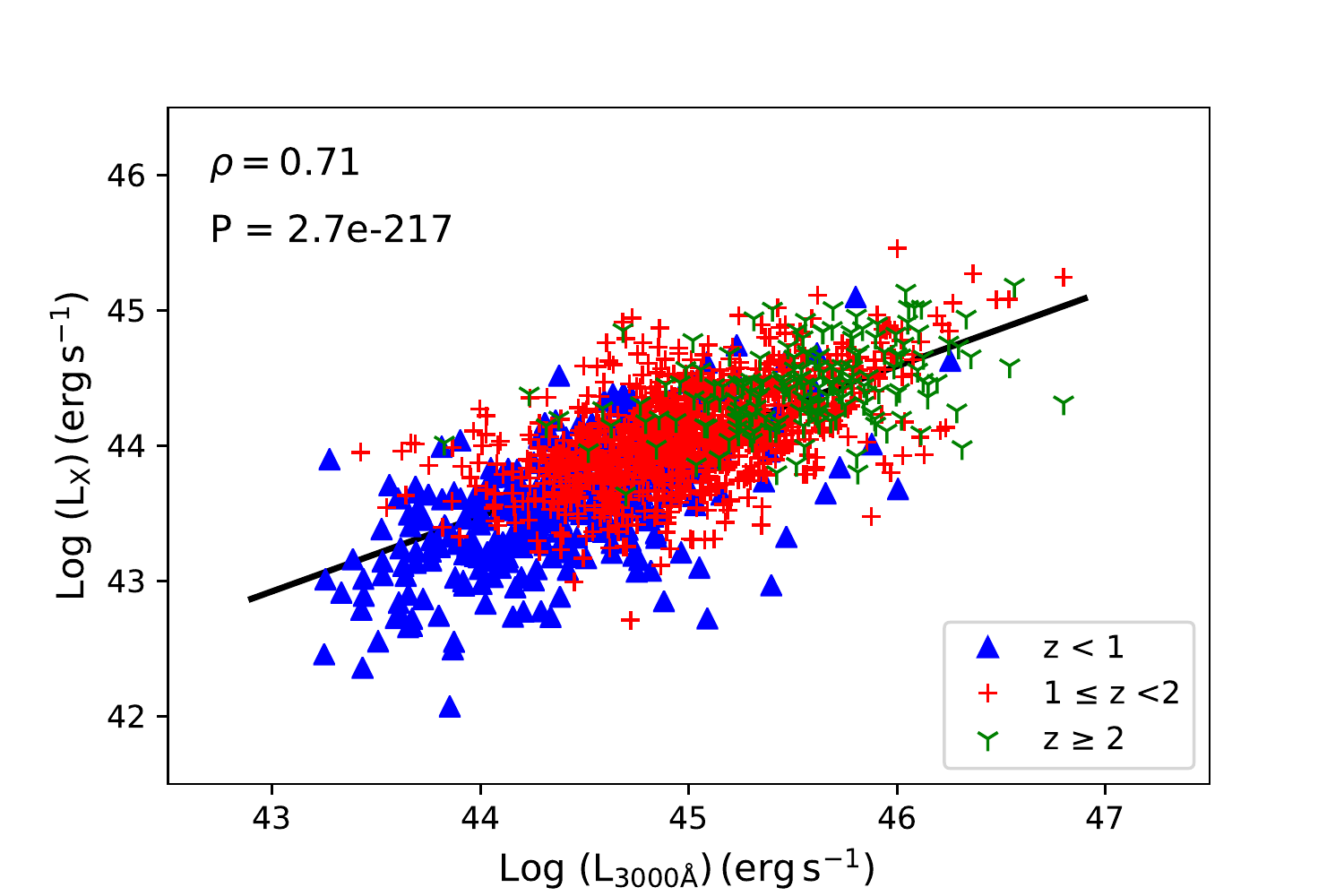}
\caption{\label{fig:8}Soft X-ray luminosity versus L$_{3000\text{\AA}}$. The solid line shows the linear best-fit for the entire sample. Sources at low, intermediate, high redshifts are shown with different markers (see the legend).}
\end{figure}
Similar to previous studies, we also noted a non-linear relation between the two luminosities.

\citet{risaliti2015hubble} stated that the non-linear relation does not show any change with respect to the redshift and therefore can be used to estimate cosmological parameters. To check if our sample follow the same trend, we divided our sample into small redshift bins with $\Delta$ log z = 0.08 (Fig. 10). Later, we performed a linear fit for the data present in each bin using the equation log L$_{\text{X}}$= a$_3$ log L$_{\text{UV}}$ + b$_3$ without constraining the fitting parameters. We ignored the first redshift bin as it contains $<$ 10 sources. In Fig. 11, we display the relation between slopes of each bin and the redshift. We too did not find any significant change in the value of the slope over the different redshift bins. The average values of slope is found to be $\overline{a}$= 0.521$\pm$0.041. Our result also supports the existence of a universal mechanism through which the energy is steadily transferred from the disc to corona causing it to radiate at higher temperature, enough to maintain the stable emission \citep{risaliti2015hubble}.
\begin{figure}
\centering
\includegraphics[width=\linewidth]{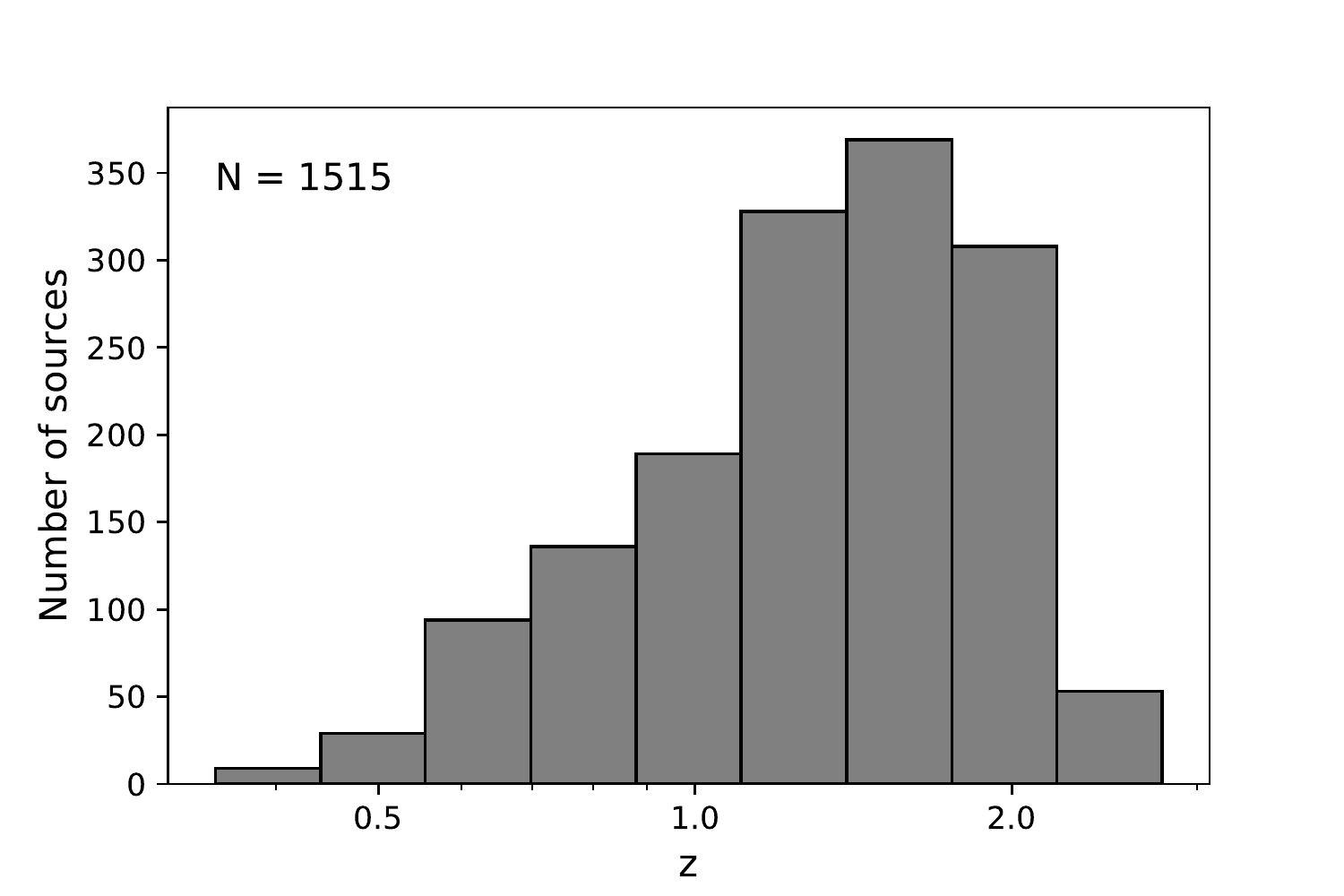}
\caption{\label{fig:9}Redshift distribution of the sample of 1515 AGNs for which both L$_{\text{X}}$ and L$_{3000\text{\AA}}$ were calculated.}
\end{figure}
\begin{figure}
\centering
\includegraphics[width=\linewidth]{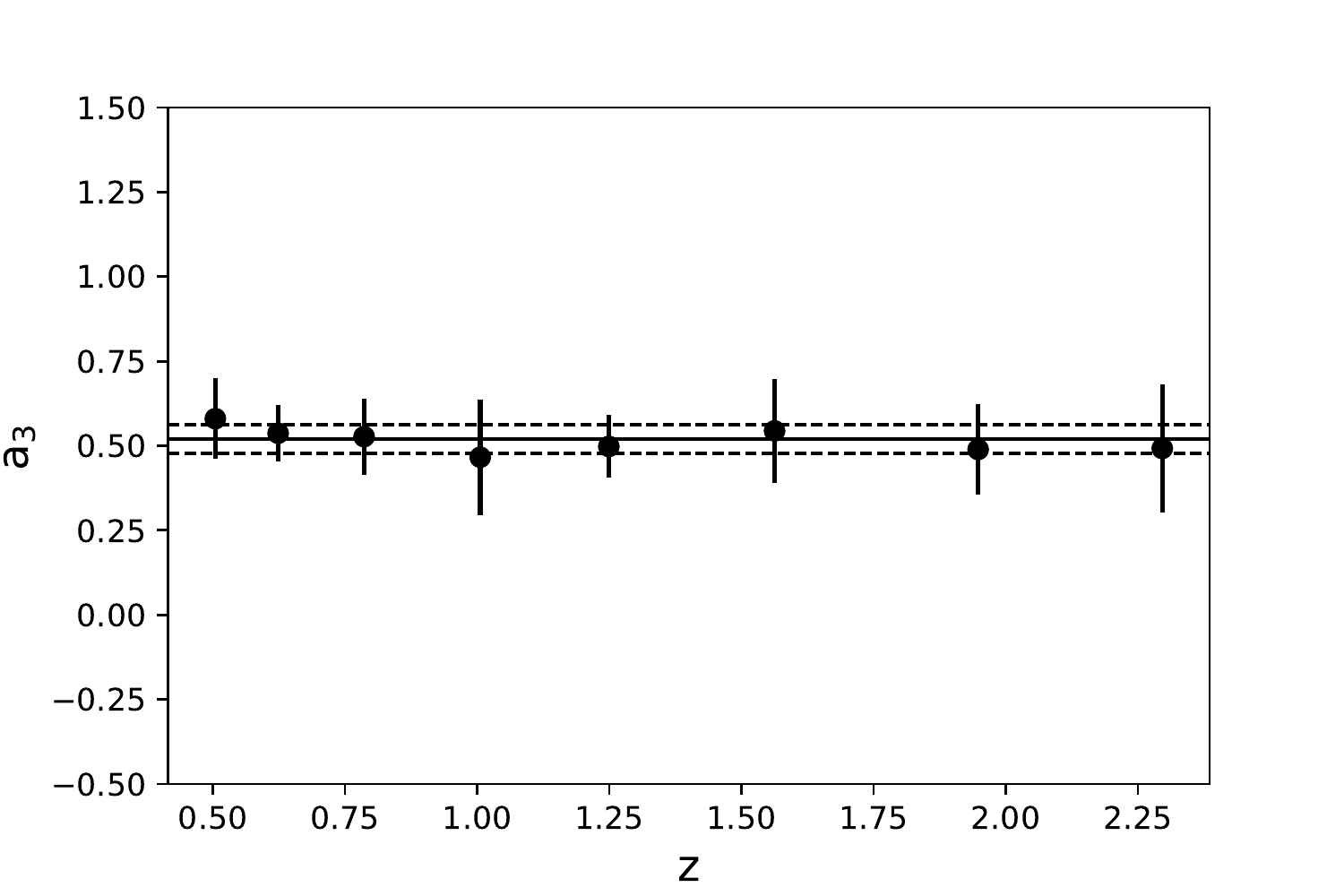}
\caption{\label{fig:10}Best-fitting values of the slopes in each redshift (z) bin. The horizontal thick line show the average value $\overline{a_3}$= 0.521$\pm$0.041, the two dashed lines indicate the $\pm$1$\sigma$ region.}
\end{figure}

\subsubsection{L$_{\text{X}}$ vs L$_{\text{MgII}}$ \& FWHM$_{\text{MgII}}$ }
Mg$\text{II}$ $\lambda$ 2800\AA\, is a UV emission line that has a broad component possibly emitted from the BLR. Similar to the relation between L$_{\text{X}}$ vs L$_{\text{H}\beta}$ discussed in section (3.2.2), we studies the connection between L$_{\text{X}}$ and L$_{\text{MgII}}$ and found L$_{\text{X}}$ to correlate firmly with the luminosity of the core component as well as the luminosity of the wings. The strongest correlation was observed for the addition of the two components; $\rho$ = 0.76 (Fig. 12) and it follows the relations based on the best-fits.
\begin{equation}
\log(\text{L}_{\text{X}})=(0.658\pm0.014) \log(\text{L}_{\text{MgII}})+(15.629\pm0.615)
\end{equation}
The above relation clearly suggests that the soft X-ray emission is affecting the UV luminosity around MgII line in the BLR. 

\begin{figure*}{}
\centering
\includegraphics[width=\linewidth]{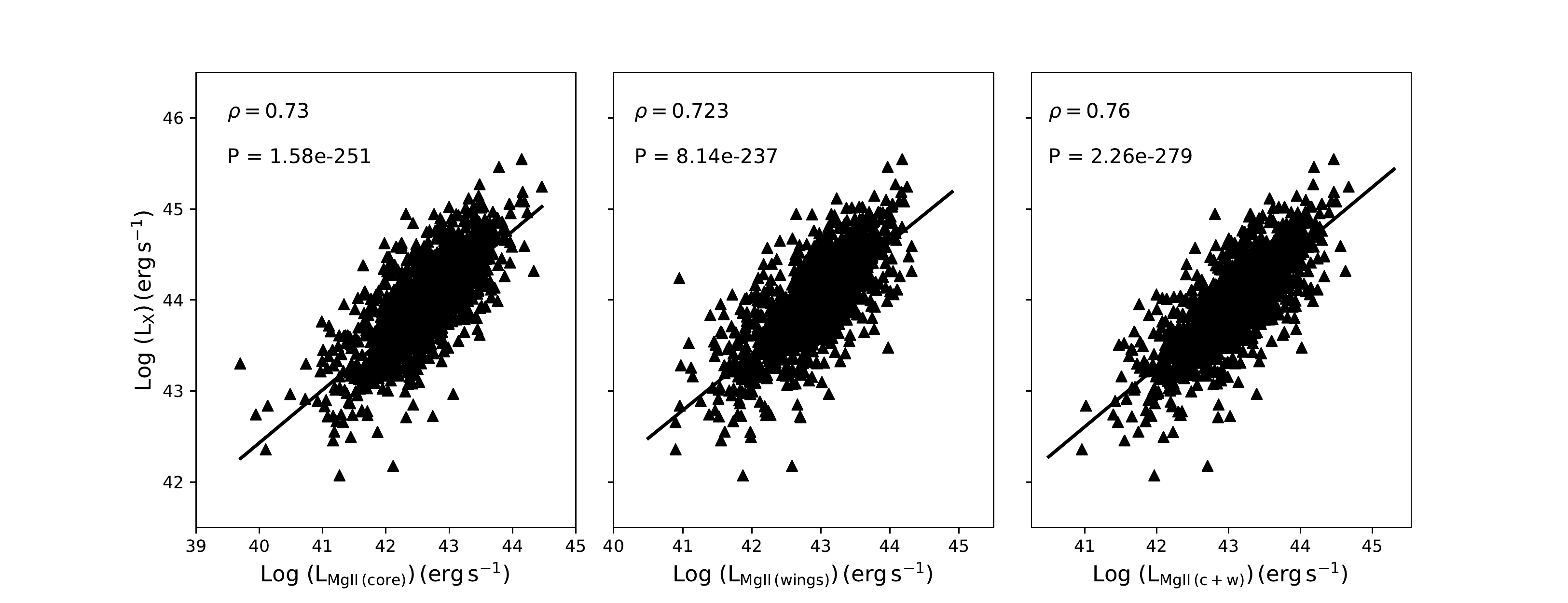}
\caption{\label{fig:11}Correlation between soft X-ray luminosity and luminosity of MgII line components (i.e. core, wings, and core+wings, from left to right panels) along with best-fits shown in lines.}

\end{figure*}

Since the MgII line is a manifestation of various components affecting the width of the line, we used a linear fit to further probe the dependency of the L$_{\text{X}}$ on the FWHM of MgII. Unlike FWHM of H${\beta}$ line, we observed a considerable correlation between L$_{\text{X}}$ and FWHM$_{\text{MgII}}$ where $\rho$ = 0.37 for the broad component (core+wings) (Fig. 13) 
\begin{equation}
\log(\text{L}_{\text{X}})=(1.081\pm0.071) \log(\text{FWHM}_ {\text{MgII}})+(40.070\pm0.259)
\end{equation}

The slope of this equation is well in agreement with the relation found by \citet{lusso2017quasars} for a sample of 545 AGNs (L$_{\text{2keV}}$ $\propto$ FWHM $^{0.971}_{\text{MgII}}$).
 We also fitted MgII line with two components, one for the core and the other for wings and further to check the origin of this correlation we studied the relation between L$_{\text{X}}$ and each component individually. The core component of MgII line did not show any correlation with L$_{\text{X}}$ contrary to the FWHM of the wings component that exhibits a relatively strong correlation $\rho$ = 0.45.
\begin{figure*}{}
\centering
\includegraphics[width=\textwidth]{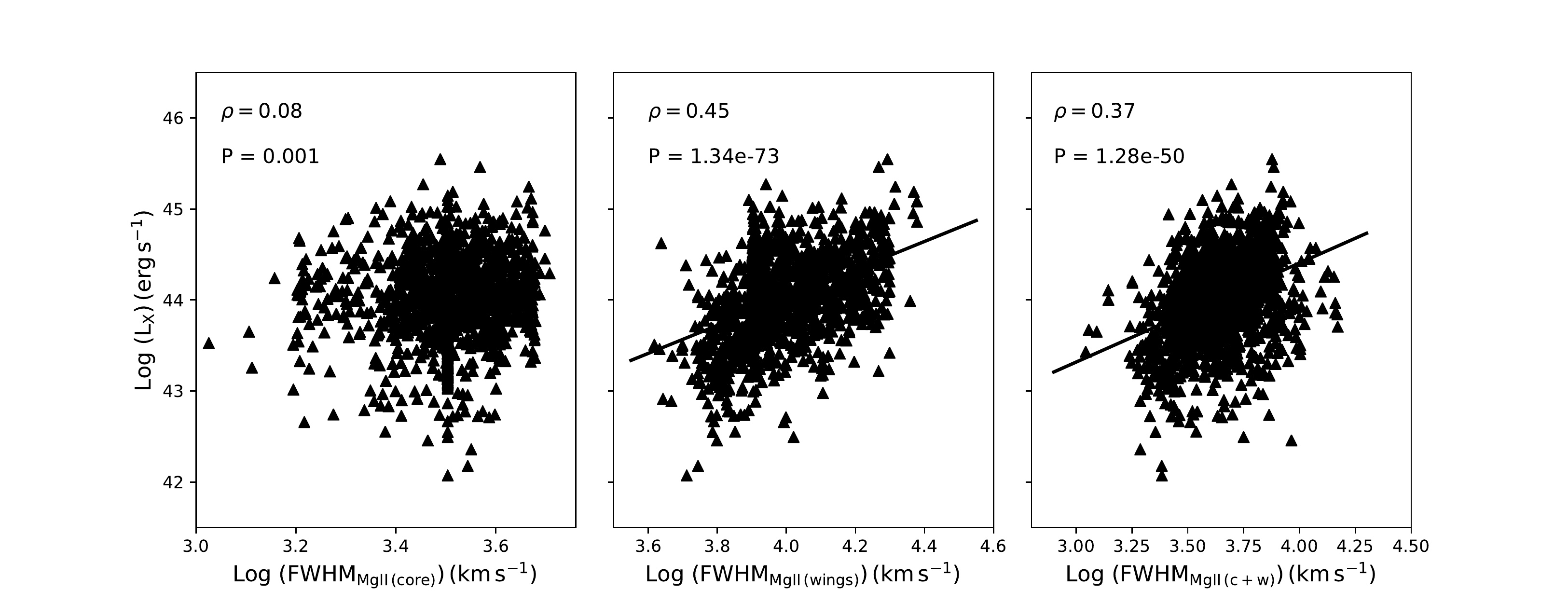}
\caption{\label{fig:12}Correlation between soft X-ray luminosity and FWHMs of MgII line components (i.e. core, wings, and core+wings, from left to right panel). Lines in second and third panels display the linear best-fits.}
\end{figure*}
Many studies has been carried out to explain the geometry of BLR based on the relations obtained between widths of different broad emission lines. In \citet{goad2012broad} model, the BLR has a bowl-shaped geometry and the broadening of emission lines at larger radius is mainly due to the turbulence that causes velocity to increase with increasing height. \citet{leon2013flare} suggested that MgII emission is connected with outflows, this was confirmed later by other studies \citep{jonic2016virilization,popovic2019structure,savic2020estimating}, that considered the wings of MgII emission lines are caused by inflow/outflow in the BLR. The observed strong correlation between of FWHM$_{\text{MgII}}$ of wings component and L$_{\text{X}}$ and its absence in case of the core component is a clear evidence that the core of MgII line is virialized whereas the wings are of different origin that might be due to outflow/inflow in BLR. Assuming that the soft excess is due to the reflection of hard X-rays then it can be concluded that the reflection in the disc plays a key role in launching outflow or inflow in the BLR. Overall the core component of MgII is arising from the BLR whereas the outflow/inflow phenomena are occurring over the disc region as they are more tightly connected to the L$_{\text{X}}$.

This also explains why we did not observe a similar correlation in case of H$\beta$ line as this line is totally virialized and soft excess does not influences the width of the line.
In order to strengthen the relation obtained between L$_{\text{X}}$ and L$_{\text{UV}}$ and test its connection with the broad line region we included FWHM$_{\text{MgII}}$ to the equation. We limited our choice to sources that have FWHM $>$ 2000 km s$^{-1}$. As shown in Fig. 14 we plotted the relation between L$_{\text{X}}$, L$_{\text{UV}}$, and FWHM$_{\text{MgII}}$. The best-fitting plane is given in equation 11
\begin{multline}
    \log(\text{L}_{\text{X}})=(0.520\pm0.014) \log(\text{L}_{\text{3000\AA}})\\
    +(0.525\pm0.053)\log(\text{FWHM}_{\text{MgII}})+(8.710\pm0.646)
\end{multline}
The slopes values are in good agreement with \citet{lusso2017quasars} theoretical model (L$_{\text{X}}$ $\propto$ L $^{4/7}_{\text{UV}}$ FWHM $^{4/7}$).
\begin{figure}{}
\begin{subfigure}{\linewidth}
\centering
\includegraphics[width=\linewidth]{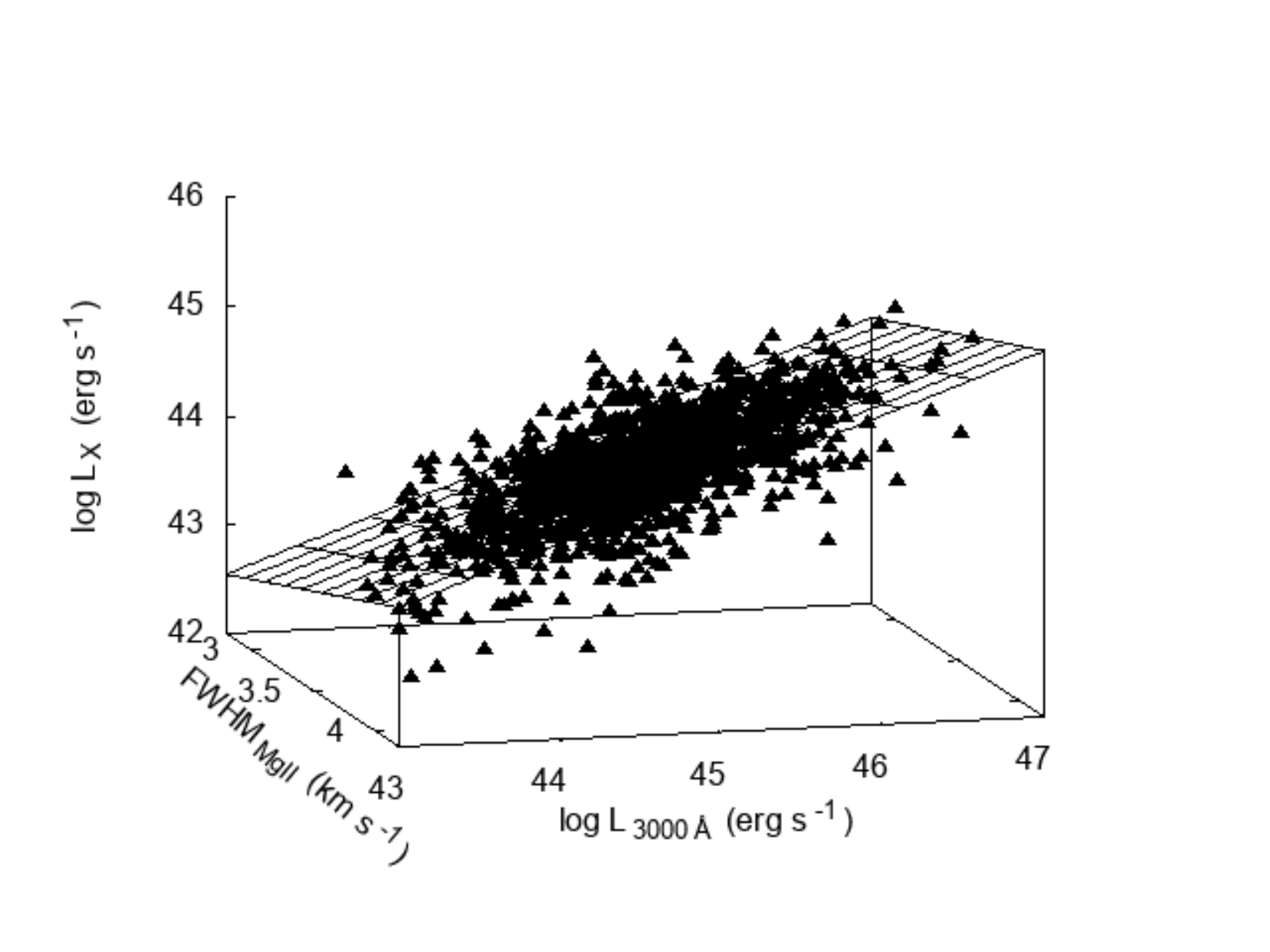}
\end{subfigure}
\newline
\begin{subfigure}{\linewidth}
\centering
\includegraphics[width=\linewidth]{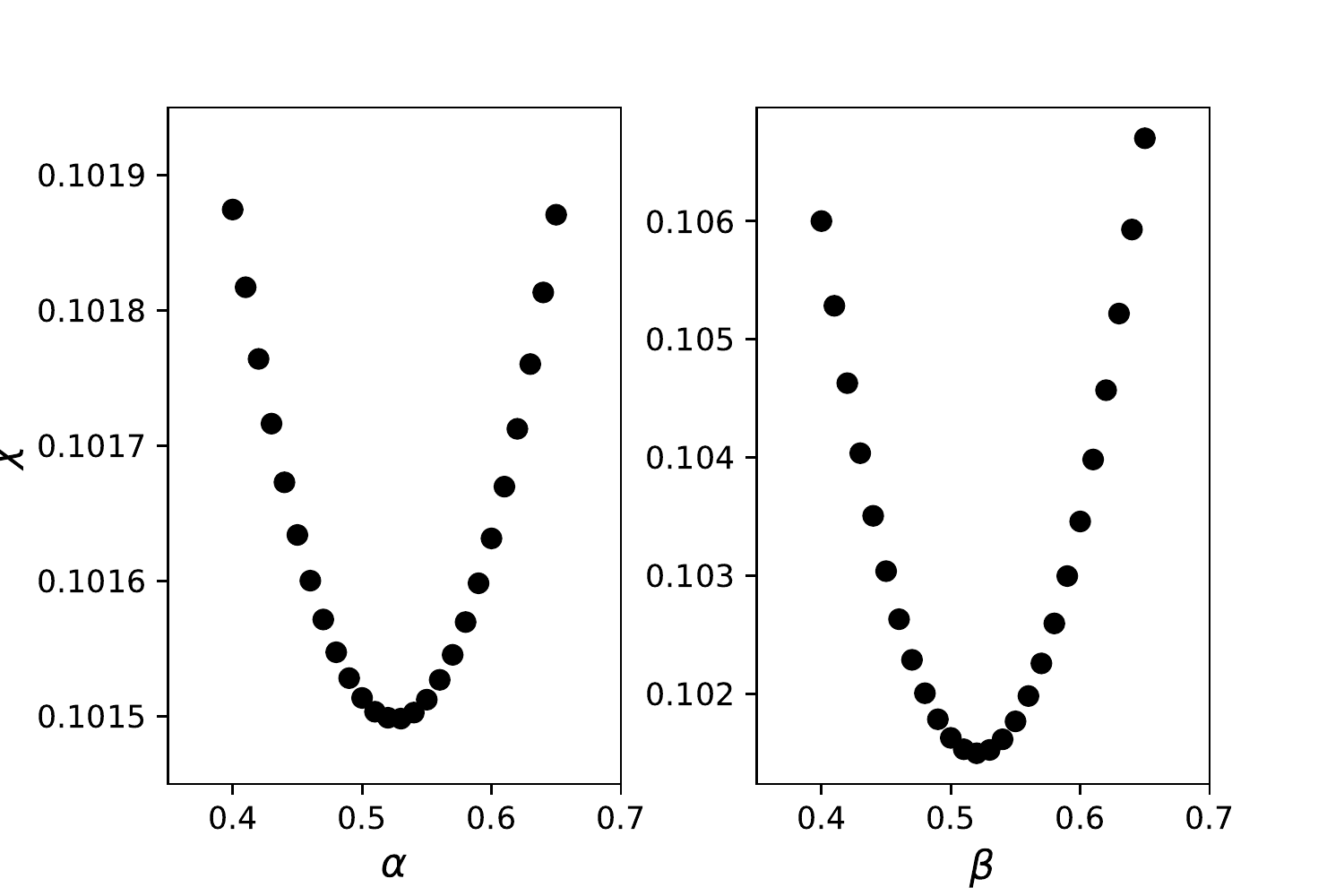}
\end{subfigure}
\caption{Upper panel: L$_{\text{X}}$ - L$_{\text{UV}}$-FWHM$_{\text{MgII}}$ relation for a sample of 1492 AGNs pertaining FWHM$_{\text{MgII}}$ $>$ 2000 km s$^{-1}$ and the plane in the figure represents the best-fit for the data.
Lower panels: $\upchi^2$ versus the values of slopes $\alpha$ and $\beta$.}
\label{fig:fig}
\end{figure}
In this toy model they connected the accretion power emitted by corona with the BLR size, they assumed the coronal emission is powered by accretion disc at transition radius where gas pressure equates the radiation pressure. In order to check the robustness of the estimated values of slopes, we changed the values of a and b between 0.4--0.6 and found the lowest values of $\upchi^2$ to be at a=0.52, b=0.52 (Fig. 14), these values are in agreement with the slopes values reported in equation (11).
Similar to \citet{lusso2017quasars} we tested the evolution of L$_\text{X}$--L$_\text{UV}$--FWHM relation with time by following the same steps explained in 3.3.1. For sources exist in each redshift bin of Fig. 15, we fitted an equation of the form log L$_{\text{X}}$ = $\alpha$ log L$_{\text{UV}}$ + $\beta$ log FWHM$_{\text{MgII}}$ + $\gamma$, we found the mean values of a and b to be 0.453 and 0.468 respectively. These two values did not exhibited any changes with z as shown in Fig. 16, therefore this relation too can be used to estimate the cosmological constants.
\begin{figure}{}
\centering
\includegraphics[width=\linewidth]{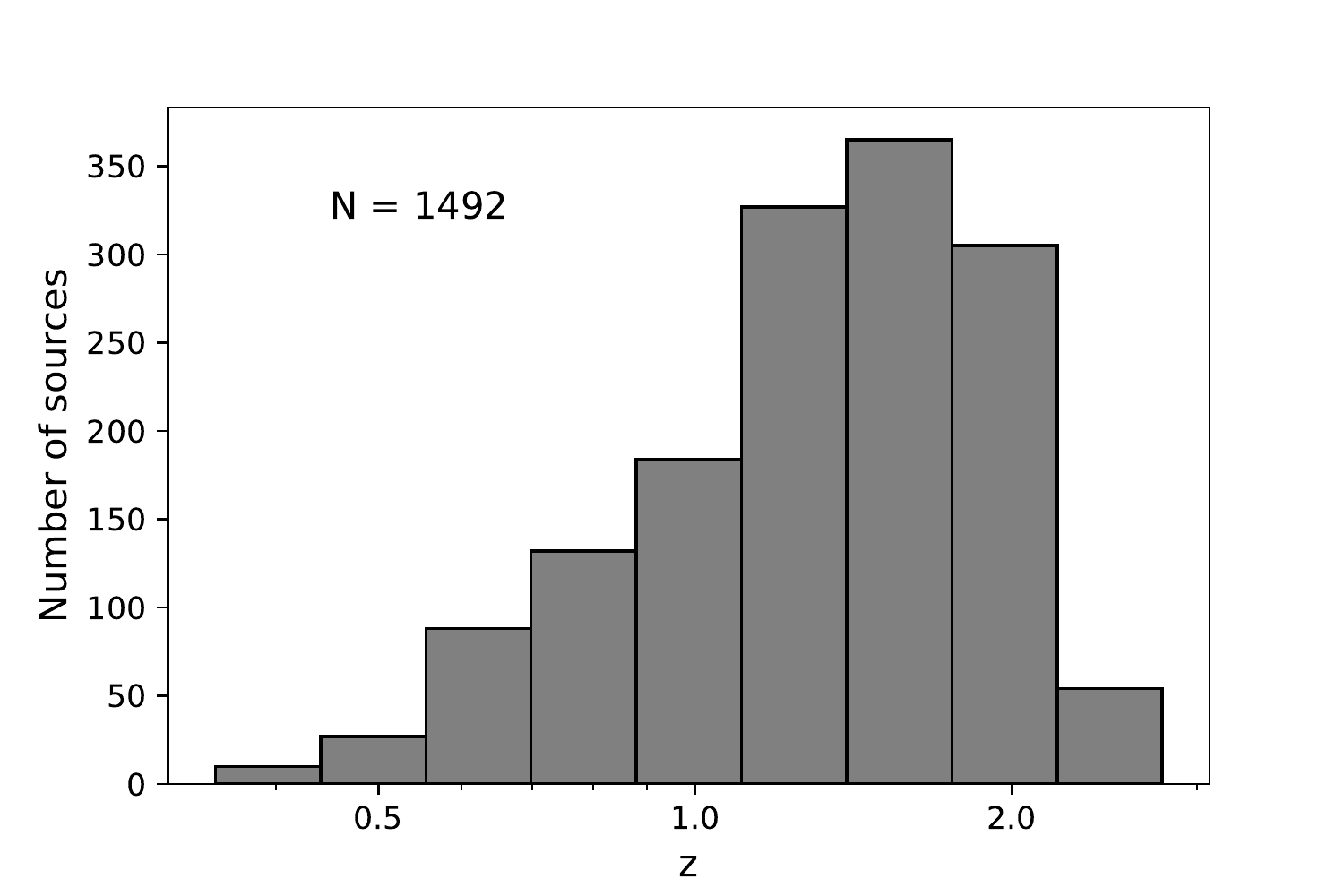}
\caption{\label{fig:14}Redshift distribution of the sample of 1492 AGNs that have L$_{\text{X}}$, L$_{3000\text{\AA}}$ and FWHM of MgII $>$2000 km s$^{-1}$.}
\end{figure}

\begin{figure}{}
\centering
\includegraphics[width=\linewidth]{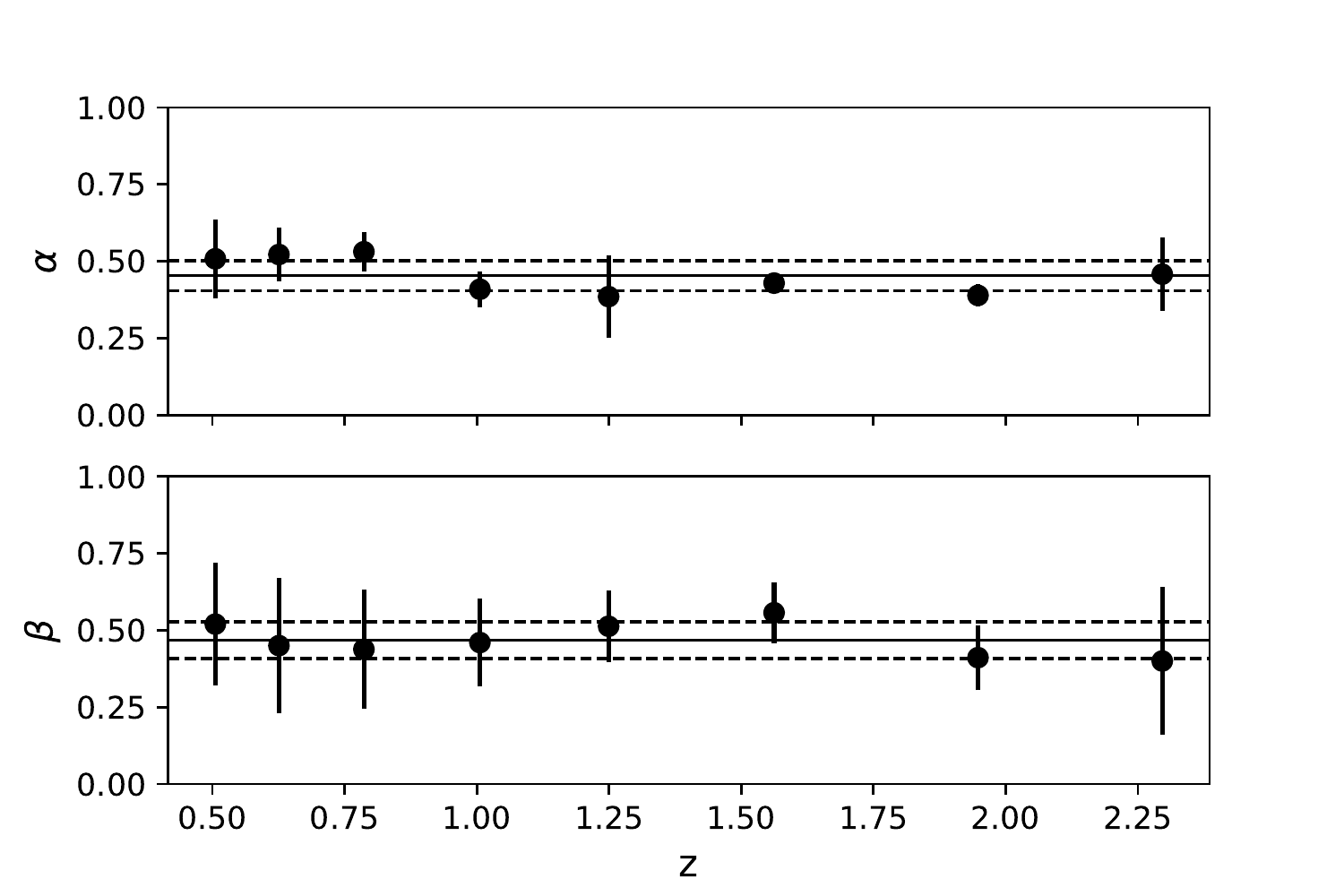}
\caption{\label{fig:15}Slopes of L$_{\text{X}}$ - L$_{\text{UV}}$-FWHM$_{\text{MgII}}$ plane as a function of redshift. The horizontal lines show the average value $\overline{\alpha}$ = 0.453$\pm$0.049 (top), $\overline{\beta}$ = 0.468$\pm$0.061 (bottom) along with $\pm$1$\sigma$ region represented by dashed lines.}
\end{figure}

\section{Results and conclusion}
For a sample of 1811 Type I AGNs extracted from SDSS database, we analyzed their optical spectra and estimated the parameters of emission lines existing in the optical/UV  domain of the spectrum to understand the relation between them and the effect of soft X-ray excess on these parameters.\\
\begin{enumerate}
 \item The optical (H$\beta$) and UV (MgII) emission lines of the BLR correlates strongly in luminosity as well as FWHM. We found MgII line to be narrower than H$\beta$ and hence probably originating from a region that extended to a larger radius than H$\beta$ emitting region. Moreover the wings of MgII line needs to arise from a different region as discussed by \citet{popovic2019structure}. We found no correlation between L$_{X}$ (0.5-2.0 keV) and FWHM$_{H\beta}$ components. Since not much work has been done, future studies would be helpful to confirm this result.

\item It is found that the relation between L$_{X}$ and  L$_{3000\text{\AA}}$ or L$_{5100\text{\AA}}$ suggests that a similar mechanism is at work in linking soft X-ray and UV or optical luminosities in simple scenario as suggested by \citet{lusso2017quasars}. However if the soft excess is due to partial covering scenario, where the primary X-rays emitted by the corona are reprocessed in an absorbing material with varying densities and ionization states assumed to be located close the corona, this will affect the correlation \citep[e.g.][]{tanaka2004partial}. In this model, the dependency of L$_{X}$ -- L$_{3000\text{\AA}}$ or L$_{5100\text{\AA}}$ may vary depending on the varying flux levels especially in the X-ray band. If we assume that the soft excess is a result of reflection phenomenon then the soft X-rays are strongly affected by the light bending effect \citep{miniutti2004light} near the SMBH unrelated to the UV and optical emission. In this scenario also the correlation varies, more detailed studies are necessary to predict the correlation.


\item We confirm the non-linearity observed in the relation between L$_{\text{X}}$ and L$_{\text{UV}}$, even in case of soft X-ray range (0.5-2 keV). This relation is not changing over redshift even after including the FWHM to the correlation and hence can be used to estimate the cosmological parameters. However, in general the uncertainties in the correlation is affected by the absorption in UV and X-ray energy bands, contribution of dust, observational contaminants in the UV and Eddington bias \citep{risaliti2019cosmological}. Furthermore the intrinsic variation of X-ray flux in AGNs will also affect the observed correlation. Future observations are necessary to precisely constrain the observed degree of non-linearity between L$_{\text{X}}$ --L$_{\text{UV}}$ relation.

\item Soft X-ray luminosity has no effect on the kinematics of the BLR emitting H$\beta$ line and MgII line core. Whereas it plays a key role in the outflow/inflow that causes the broadening of MgII line and responsible for its wings. If we assume that the soft excess is produced due to the reflection scenario where a fraction of hard X-rays emitted by corona are incident and subsequently reflected in the disc, then this mechanism is important in order to understand the phenomenon of outflow/inflows in AGNs.

\end{enumerate}

\begin{table*}
\caption{Spectral parameters for the first 75 objects of the 1811 AGNs sample are displayed. Full version of the table is available as a supplementary material. Column 3: redshift; column 4: soft X-ray luminosity in (0.5--2.0 keV) calculated using fluxes given by \citet{liu2016x}; column 5: continuum luminosity at 3000\AA; column 6: continuum luminosity at 5100\AA; columns 7,8,9 and 10 : FWHMs of H$\beta$ components; columns 11,12, 13 and 14 : luminosities of H$\beta$ components; columns 15, 16,and 17: FWHMs of MgII components; columns 18, 19, and 20: luminosities of MgII components.}
\label{tab:T1}
\setlength\tabcolsep{3 pt}
\setlength{\cmidrulekern}{2em}
\scriptsize
\centering
\begin{tabular}{|l|c|c|c|c|c|cccc|cccc|ccc|ccc|}

\hline
 &  &  & Log (L$_{\text{X}}$) & Log(L$_{3000\text{\AA}}$) & Log(L$_{5100\text{\AA}}$)& \multicolumn{4}{c|}{Log(FWHM H$\beta$) (km s$^{-1}$)} &\multicolumn{4}{c|}{ Log(L$_{\text{H}\beta}$) (erg s$^{-1}$) }  & \multicolumn{3}{c|}{Log(FWHM MgII)(km s$^{-1}$)} & \multicolumn{3}{c|}{Log(L MgII) (erg s$^{-1}$)}\\
 \cline{7-10} \cline{11-14} \cline{15-17} \cline{18-20} 
ID & SDSS name & z & (erg s$^{-1}$)& (erg s$^{-1}$)& (erg s$^{-1}$)&
 NLR & ILR & VBLR & BLR & NLR & ILR & VBLR & BLR & core & wings & c+w& core & wings & c+w\\
 (1) & (2) & (3) & (4) & (5) & (6) & (7) & (8) & (9) & (10) & (11) & (12) & (13) & (14) & (15) & (16) & (17) & (18)& (19)& (20)\\

\hline
1 & 020550.94-065350.0 & 0.59 & 42.46 & 43.25 & -- & 2.50 & 3.16 & 3.76 & 3.40 & 40.52 & 41.04 & 41.64 & 41.74 & 3.30 & 4.10 & 3.16 & 41.08 & 41.55 & 41.55 \\
2 & 023025.82-042354.5 & 0.60 & 43.01 & 43.26 & -- & -- & -- & -- & -- & -- & -- & -- & -- & 3.41 & 3.78 & 3.61 & 41.35 & 41.83 & 41.95 \\
3 & 022149.33-052454.9 & 0.66 & 43.90 & 43.28 & -- & 2.40 & 3.64 & 3.94 & 3.79 & 41.03 & 41.15 & 41.45 & 41.63 & 3.30 & 4.01 & 3.41 & 41.51 & 41.84 & 42.00 \\
4 & 022415.76-052720.0 & 0.56 & 42.91 & 43.33 & 42.51 & 2.40 & 3.03 & 3.72 & 3.24 & 41.26 & 40.87 & 41.65 & 41.72 & 3.14 & 4.07 & 3.18 & 40.73 & 41.52 & 41.58 \\
5 & 022116.34-050819.0 & 0.54 & 43.16 & 43.39 & 43.28 & 2.34 & 3.05 & 3.62 & 3.28 & 40.79 & 40.82 & 41.40 & 41.50 & 3.42 & 4.08 & 3.47 & 41.27 & 41.14 & 41.51 \\
6 & 020334.58-051721.3 & 1.13 & 43.95 & 43.43 & -- & -- & -- & -- & -- & -- & -- & -- & -- & 3.40 & 3.70 & 3.72 & 41.34 & 41.55 & 41.76 \\
7 & 022205.99-032346.4 & 0.57 & 42.79 & 43.43 & 43.37 & 2.29 & 3.27 & 3.72 & 3.33 & 39.77 & 40.98 & 41.06 & 41.32 & 3.39 & 3.84 & 3.44 & 41.62 & 41.47 & 41.85 \\
8 & 020655.39-040805.1 & 0.55 & 42.36 & 43.43 & 43.44 & 2.58 & 3.01 & 3.97 & 3.51 & 40.94 & 40.82 & 41.82 & 41.86 & 3.08 & 3.93 & 3.29 & 40.10 & 40.90 & 40.96 \\
9 & 021936.57-033358.8 & 0.72 & 43.01 & 43.44 & 44.22 & 2.45 & 3.16 & 3.76 & 3.68 & 41.13 & 42.67 & 42.86 & 41.89 & 3.34 & 3.69 & 3.54 & 41.21 & 41.52 & 41.69 \\
10 & 021037.14-061614.9 & 0.53 & 42.89 & 43.44 & 43.51 & 2.69 & 3.16 & 3.76 & 3.31 & 40.27 & 41.15 & 41.53 & 41.68 & 3.17 & 3.86 & 3.29 & 40.92 & 41.26 & 41.42 \\
11 & 021950.50-040141.0 & 0.51 & 42.55 & 43.51 & 43.37 & 2.41 & 3.10 & 3.84 & 3.21 & 39.87 & 41.05 & 41.39 & 41.55 & 3.35 & 3.88 & 3.54 & 41.19 & 41.61 & 41.75 \\
12 & 020655.10-064551.6 & 0.56 & 43.38 & 43.53 & 43.64 & 2.68 & 3.16 & 3.77 & 3.40 & 40.71 & 40.97 & 41.58 & 41.67 & 3.30 & 4.06 & 3.43 & 41.10 & 41.46 & 41.62 \\
13 & 020734.47-060542.7 & 0.65 & 43.14 & 43.53 & 43.32 & 2.49 & 3.16 & 3.76 & 3.24 & 32.39 & 41.38 & 41.44 & 41.71 & 3.31 & 4.02 & 3.45 & 41.50 & 41.92 & 42.06 \\
14 & 020813.75-043651.5 & 0.59 & 43.04 & 43.53 & 43.45 & 2.52 & 3.17 & 3.92 & 3.35 & 39.67 & 40.90 & 41.52 & 41.61 & 3.31 & 4.03 & 3.43 & 41.19 & 41.54 & 41.70 \\
15 & 021537.17-045657.2 & 1.03 & 43.54 & 43.55 & -- & -- & -- & -- & -- & -- & -- & -- & -- & 3.40 & 3.71 & 3.50 & 41.49 & 41.45 & 41.77 \\
16 & 023049.60-052000.1 & 0.77 & 43.71 & 43.56 & 43.24 & 2.29 & 3.16 & 3.76 & 3.76 & 40.27 & 41.49 & 41.96 & 41.96 & 3.42 & 4.05 & 3.50 & 41.89 & 41.91 & 42.20 \\
17 & 021448.84-040601.7 & 0.44 & 42.73 & 43.59 & 43.15 & 2.67 & 3.16 & 3.77 & 3.47 & 40.06 & 40.96 & 41.68 & 41.75 & 3.35 & 4.00 & 3.63 & 41.17 & 41.85 & 41.93 \\
18 & 022509.79-051246.5 & 0.87 & 43.61 & 43.60 & 44.07 & 2.37 & 3.14 & 3.96 & 3.94 & 37.33 & 41.28 & 42.78 & 42.78 & 2.96 & 3.86 & 3.72 & 41.26 & 42.51 & 42.53 \\
19 & 021335.38-043941.3 & 0.70 & 42.84 & 43.61 & 43.86 & 2.63 & 3.16 & 3.77 & 3.77 & 40.75 & 41.33 & 41.96 & 41.96 & 2.96 & 3.70 & 3.44 & 40.13 & 40.96 & 41.02 \\
20 & 020542.94-065941.3 & 0.69 & 43.24 & 43.62 & 43.96 & 2.32 & 3.12 & 3.73 & 3.73 & 40.32 & 41.99 & 41.87 & 41.87 & 3.42 & 4.03 & 3.68 & 41.48 & 42.09 & 42.18 \\
21 & 021657.09-053159.5 & 0.46 & 42.74 & 43.62 & 43.89 & 2.61 & 3.54 & 3.87 & 3.78 & 41.29 & 41.14 & 41.91 & 41.98 & 3.37 & 4.15 & 3.53 & 41.29 & 41.79 & 41.91 \\
22 & 022944.55-050750.9 & 1.59 & 43.96 & 43.62 & -- & -- & -- & -- & -- & -- & -- & -- & -- & 3.40 & 4.13 & 3.92 & 42.09 & 42.74 & 43.09 \\
23 & 022441.65-045001.5 & 0.90 & 43.11 & 43.63 & 44.09 & 2.53 & 3.16 & 3.76 & 3.33 & 41.03 & 41.49 & 41.91 & 42.05 & 3.29 & 3.95 & 3.41 & 41.69 & 41.97 & 42.15 \\
24 & 023309.17-050243.3 & 0.52 & 43.04 & 43.64 & 43.29 & 2.65 & 3.16 & 3.77 & 3.33 & 40.74 & 41.12 & 41.56 & 41.70 & 3.37 & 4.04 & 3.44 & 41.27 & 41.43 & 41.66 \\
25 & 021353.83-050933.3 & 1.15 & 43.63 & 43.64 & -- & -- & -- & -- & -- & -- & -- & -- & -- & 3.41 & 3.70 & 3.48 & 41.73 & 41.57 & 41.96 \\
26 & 023103.39-050202.4 & 0.65 & 42.66 & 43.65 & 43.72 & 2.51 & 3.16 & 3.76 & 3.27 & 40.25 & 41.36 & 41.56 & 41.77 & 3.49 & 3.72 & 3.51 & 41.32 & 40.89 & 41.46 \\
27 & 022803.33-054014.8 & 0.56 & 42.90 & 43.65 & 42.97 & 2.45 & 3.16 & 3.77 & 3.31 & 40.05 & 41.20 & 41.59 & 41.74 & 3.33 & 4.00 & 3.66 & 41.06 & 41.82 & 41.89 \\
28 & 021454.15-051414.2 & 0.57 & 43.49 & 43.66 & 43.70 & 2.47 & 3.01 & 3.98 & 3.59 & 40.93 & 40.54 & 41.60 & 41.63 & 3.20 & 3.87 & 3.29 & 41.25 & 41.67 & 41.81 \\
29 & 020718.49-041629.9 & 0.76 & 43.40 & 43.66 & 43.13 & 2.56 & 3.28 & 3.85 & 3.39 & 40.95 & 41.31 & 41.56 & 41.75 & 3.35 & 4.07 & 3.55 & 41.30 & 41.86 & 41.96 \\
30 & 021421.45-035117.2 & 1.30 & 44.02 & 43.67 & -- & -- & -- & -- & -- & -- & -- & -- & -- & 3.40 & 4.04 & 3.57 & 41.56 & 41.92 & 42.08 \\
31 & 021502.99-051141.5 & 0.53 & 42.67 & 43.67 & 43.31 & 2.44 & 3.16 & 3.76 & 3.76 & 40.94 & 41.01 & 40.88 & 40.88 & 3.29 & 4.08 & 3.46 & 41.22 & 41.73 & 41.85 \\
32 & 020454.66-051014.4 & 0.70 & 42.72 & 43.67 & -- & 2.29 & 3.37 & 3.76 & 3.37 & 40.69 & 41.12 & 41.09 & 42.12 & 3.28 & 4.10 & 3.43 & 41.08 & 41.52 & 41.66 \\
33 & 020834.45-041557.5 & 1.05 & 44.01 & 43.69 & -- & -- & -- & -- & -- & -- & -- & -- & -- & 3.40 & 4.10 & 3.47 & 42.02 & 42.18 & 42.41 \\
34 & 022918.27-053804.1 & 0.48 & 43.70 & 43.69 & 43.57 & 2.37 & 3.20 & 4.01 & 3.33 & 41.35 & 41.33 & 41.78 & 41.91 & 3.28 & 4.02 & 3.39 & 41.57 & 41.90 & 42.07 \\
35 & 021909.73-032318.4 & 0.54 & 43.13 & 43.69 & 43.47 & 2.71 & 3.16 & 3.78 & 3.29 & 40.91 & 41.16 & 41.46 & 41.63 & 3.23 & 3.77 & 3.31 & 41.57 & 41.85 & 42.03 \\
36 & 021535.93-035612.9 & 0.74 & 43.21 & 43.69 & 43.64 & 2.53 & 3.16 & 3.76 & 3.56 & 40.36 & 40.69 & 41.56 & 41.61 & 3.05 & 3.77 & 3.61 & 40.97 & 41.77 & 41.83 \\
37 & 021835.91-053757.9 & 0.39 & 43.65 & 43.70 & 44.01 & 2.55 & 3.28 & 3.87 & 3.56 & 40.48 & 41.39 & 42.02 & 42.11 & 3.46 & 4.09 & 3.63 & 41.13 & 41.55 & 41.69 \\
38 & 022921.92-044219.9 & 0.78 & 43.53 & 43.70 & 44.18 & 2.29 & 3.09 & 3.79 & 3.26 & 40.01 & 41.70 & 42.19 & 42.32 & 3.48 & 3.83 & 3.52 & 41.36 & 41.09 & 41.55 \\
39 & 020703.36-062917.2 & 0.59 & 43.51 & 43.71 & 43.73 & 2.29 & 3.01 & 3.81 & 3.14 & 40.58 & 41.10 & 41.68 & 41.78 & 3.12 & 3.79 & 3.36 & 42.24 & 41.48 & 41.48 \\
40 & 022047.30-042619.0 & 0.83 & 42.86 & 43.72 & 44.10 & 2.50 & 3.16 & 3.76 & 3.28 & 32.53 & 41.32 & 41.59 & 41.78 & 3.27 & 4.04 & 3.40 & 41.43 & 41.83 & 41.98 \\
41 & 022258.90-055758.1 & 0.73 & 43.45 & 43.73 & 44.23 & 2.70 & 3.32 & 4.01 & 3.64 & 41.41 & 41.43 & 42.18 & 42.25 & 3.20 & 3.92 & 3.27 & 41.16 & 41.50 & 41.66 \\
42 & 021331.35-042849.5 & 0.74 & 43.64 & 43.75 & 43.95 & 2.59 & 3.16 & 3.76 & 3.62 & 40.00 & 40.81 & 41.81 & 41.85 & 3.32 & 4.06 & 3.47 & 41.69 & 42.13 & 42.27 \\
43 & 021423.22-053002.3 & 1.28 & 43.86 & 43.75 & -- & -- & -- & -- & -- & -- & -- & -- & -- & 3.40 & 3.84 & 3.46 & 41.99 & 41.87 & 42.23 \\
44 & 020504.45-065412.5 & 0.73 & 43.22 & 43.76 & 44.01 & 2.69 & 3.31 & 3.87 & 3.92 & 40.30 & 40.93 & 41.99 & 42.02 & 3.33 & 3.86 & 3.53 & 41.33 & 41.82 & 41.94 \\
45 & 021933.75-054920.1 & 0.69 & 43.30 & 43.76 & 44.12 & 2.36 & 3.21 & 3.87 & 3.85 & 40.92 & 40.87 & 41.81 & 41.86 & 2.96 & 3.86 & 3.86 & 41.70 & 42.27 & 42.27 \\
46 & 021705.45-053428.2 & 0.63 & 43.15 & 43.76 & 43.24 & 2.42 & 3.16 & 3.77 & 3.31 & 40.18 & 41.33 & 41.71 & 41.86 & 3.32 & 4.09 & 3.43 & 41.76 & 42.06 & 42.24 \\
47 & 020309.07-043709.7 & 0.60 & 43.28 & 43.78 & 42.90 & 2.29 & 3.01 & 3.34 & 3.03 & 40.39 & 40.88 & 40.67 & 41.09 & 3.38 & 3.70 & 3.16 & 41.17 & 41.72 & 41.83 \\
48 & 020543.26-051702.4 & 0.65 & 42.74 & 43.80 & 44.04 & 2.54 & 3.14 & 3.82 & 3.65 & 40.63 & 41.11 & 42.08 & 42.13 & 2.96 & 3.78 & 3.70 & 41.95 & 41.38 & 41.40 \\
49 & 021402.24-051317.4 & 0.64 & 43.25 & 43.80 & 43.96 & 2.29 & 3.16 & 3.77 & 3.31 & 39.93 & 41.34 & 41.72 & 41.87 & 3.27 & 3.57 & 3.18 & 41.07 & 41.91 & 41.97 \\
50 & 020915.99-051543.1 & 0.62 & 43.99 & 43.81 & 43.85 & 2.61 & 3.16 & 3.77 & 3.58 & 40.54 & 40.85 & 41.75 & 41.80 & 3.34 & 4.00 & 3.41 & 42.04 & 42.10 & 42.37 \\
51 & 022658.02-051606.5 & 0.91 & 43.60 & 43.82 & 43.38 & 2.52 & 3.16 & 3.76 & 3.68 & 40.38 & 40.72 & 41.94 & 41.97 & 3.40 & 4.05 & 3.89 & 41.34 & 42.17 & 42.23 \\
52 & 021550.87-043140.3 & 1.18 & 43.40 & 43.82 & -- & -- & -- & -- & -- & -- & -- & -- & -- & 3.40 & 4.06 & 3.52 & 41.63 & 41.91 & 42.09 \\
53 & 021109.34-043038.0 & 2.05 & 44.02 & 43.83 & -- & -- & -- & -- & -- & -- & -- & -- & -- & 3.41 & 4.03 & 3.75 & 41.89 & 42.65 & 42.72 \\
54 & 023427.71-045706.6 & 0.60 & 43.41 & 43.83 & 43.39 & 2.53 & 3.16 & 3.76 & 3.31 & 40.16 & 41.27 & 41.65 & 41.80 & 3.48 & 3.72 & 3.66 & 41.53 & 41.94 & 42.08 \\
55 & 020737.37-051353.2 & 0.99 & 43.40 & 43.83 & 44.12 & 2.55 & 3.16 & 3.76 & 3.76 & 41.00 & 41.83 & 41.62 & 41.62 & 3.40 & 4.07 & 4.00 & 41.95 & 42.12 & 42.12 \\
56 & 021915.89-050503.1 & 0.38 & 42.07 & 43.85 & 43.44 & 2.59 & 3.21 & 3.91 & 3.34 & 40.72 & 41.32 & 41.71 & 41.86 & 3.21 & 4.02 & 3.38 & 41.27 & 41.87 & 41.97 \\
57 & 022439.73-042401.6 & 0.48 & 43.33 & 43.86 & 44.14 & 2.66 & 3.25 & 3.86 & 3.44 & 41.31 & 41.28 & 41.83 & 41.94 & 3.55 & 4.05 & 3.19 & 40.99 & 41.76 & 41.83 \\
58 & 022608.03-030556.8 & 0.78 & 43.34 & 43.86 & 44.33 & 2.37 & 3.64 & 3.84 & 3.85 & 41.12 & 40.91 & 41.72 & 41.94 & -- & -- & -- & -- & -- & -- \\
59 & 021539.84-044639.0 & 1.56 & 43.59 & 43.86 & -- & -- & -- & -- & -- & -- & -- & -- & -- & 3.51 & 4.04 & 3.68 & 42.26 & 42.91 & 43.23 \\
60 & 022055.25-031215.2 & 1.22 & 43.99 & 43.86 & -- & -- & -- & -- & -- & -- & -- & -- & -- & 3.72 & 4.16 & 3.76 & 42.53 & 42.18 & 42.69 \\
61 & 021702.14-035645.7 & 0.63 & 42.49 & 43.87 & 43.90 & 2.38 & 3.27 & 3.72 & 3.68 & 31.64 & 40.97 & 41.79 & 41.85 & 3.52 & 4.04 & 3.75 & 41.45 & 41.99 & 42.10 \\
62 & 021006.85-051538.5 & 0.54 & 42.55 & 43.87 & 43.54 & 2.51 & 3.16 & 3.76 & 3.67 & 39.67 & 40.56 & 41.73 & 41.75 & 3.29 & 3.87 & 3.36 & 41.87 & 41.98 & 42.23 \\
63 & 020836.23-064949.0 & 0.91 & 43.65 & 43.87 & 44.12 & 2.30 & 3.17 & 3.74 & 3.18 & 39.96 & 41.76 & 40.92 & 41.82 & 3.53 & 3.67 & 3.18 & 41.83 & 42.40 & 42.50 \\
64 & 021946.98-055525.1 & 0.55 & 43.02 & 43.88 & 43.61 & 2.64 & 3.01 & 3.98 & 3.15 & 40.59 & 41.03 & 41.71 & 41.80 & 3.33 & 4.03 & 3.41 & 41.59 & 41.74 & 41.97 \\
65 & 023435.43-045431.9 & 0.63 & 43.27 & 43.89 & 43.53 & 2.71 & 3.16 & 3.77 & 3.66 & 40.16 & 40.62 & 41.74 & 41.77 & 3.42 & 4.02 & 3.59 & 41.56 & 41.99 & 42.13 \\
66 & 020450.76-041628.6 & 0.58 & 43.29 & 43.89 & 43.66 & 2.29 & 3.20 & 3.77 & 3.36 & 40.23 & 41.38 & 41.77 & 41.92 & 3.31 & 4.02 & 3.40 & 41.67 & 41.84 & 42.07 \\
67 & 020732.28-042636.7 & 1.11 & 43.33 & 43.90 & -- & -- & -- & -- & -- & -- & -- & -- & -- & 3.40 & 3.71 & 3.67 & 41.23 & 42.25 & 42.29 \\
68 & 021031.98-060515.2 & 0.97 & 44.04 & 43.90 & 44.11 & 2.29 & 3.16 & 3.77 & 3.77 & 32.79 & 41.78 & 42.46 & 42.46 & 3.36 & 4.09 & 3.50 & 42.23 & 42.59 & 42.74 \\
69 & 022548.06-051706.2 & 0.76 & 43.61 & 43.90 & -- & -- & -- & -- & -- & -- & -- & -- & -- & 3.40 & 4.01 & 3.61 & 41.36 & 41.94 & 42.04 \\
70 & 022249.13-044600.9 & 1.22 & 43.85 & 43.91 & -- & -- & -- & -- & -- & -- & -- & -- & -- & 3.55 & 3.78 & 3.78 & 42.03 & 42.09 & 42.36 \\
71 & 021946.18-050041.2 & 0.77 & 43.00 & 43.92 & 42.39 & 2.61 & 3.15 & 3.77 & 3.31 & 40.55 & 41.53 & 41.93 & 42.07 & 3.33 & 3.80 & 3.47 & 41.52 & 41.86 & 42.02 \\
72 & 022001.26-034830.0 & 0.58 & 42.96 & 43.92 & 43.71 & 2.48 & 3.17 & 3.51 & 3.62 & 41.10 & 40.91 & 41.54 & 41.63 & 3.30 & 3.87 & 3.84 & 40.49 & 41.98 & 41.99 \\
73 & 020709.83-042501.4 & 0.73 & 43.20 & 43.92 & 43.97 & 2.68 & 3.16 & 3.77 & 3.77 & 41.20 & 41.04 & 41.81 & 41.81 & 3.31 & 4.10 & 3.39 & 41.82 & 41.96 & 42.20 \\
74 & 020202.44-042819.5 & 0.72 & 43.46 & 43.92 & 43.73 & 2.59 & 3.01 & 3.90 & 3.20 & 40.77 & 41.31 & 42.06 & 42.13 & 3.13 & 3.69 & 3.20 & 42.10 & 41.65 & 41.65 \\
75 & 023128.99-044847.1 & 0.65 & 43.26 & 43.94 & 43.69 & 2.33 & 3.01 & 3.34 & 3.08 & 40.75 & 41.19 & 41.10 & 41.45 & 3.34 & 3.61 & 3.36 & 41.66 & 41.12 & 41.77 \\

\hline
\end{tabular}
\end{table*}

\begin{landscape}
\begin{table}
\caption{Correlations for optical, UV, and X-ray parameters. We reported here only correlations with P < 0.05.}
\label{tab:landscape}
\setlength\tabcolsep{3pt}
\setlength{\cmidrulekern}{0.5em}
    \centering 

\scriptsize
\begin{tabular}{lccccccccccccccccccc}
\hline
\\
 & & L$_{\text{X}}$ & $\uplambda_{\text{Edd}}$ &  z & M$_{\text{\text{BH}}}$ & L$_{3000\text{\AA}}$ & L$_{5000\text{\AA}}$ & EW H$\beta$  & EW H$\beta$ 
 & FWHM H$\beta$  & FWHM H$\beta$  & L H$\beta$ & L H$\beta$ & EW MgII & EW MgII & FWHM MgII &  FWHM MgII & L MgII & L MgII\\
 & & & & & & & & ILR & VBLR  & ILR & VBLR & ILR & VBLR & core & wings & core & wings & core & wings

 \\
 \hline
\\
L$_{\text{X}}$ & $\rho$ & 1 & 0.18 & 0.71 & 0.49 & 0.71 & 0.63 & X & 0.21 & X & X & 0.63 & 0.71 & X & X & X & 0.45 & 0.73 & 0.72 \\
 & P & 0 & 3.63E-15 & 6.003E-286 & 1.61E-112 & 2.7E-217 & 3.21E-52 & X & 2.71E-06 & X & X & 1.53E-53 & 6.23E-76 & X & X & X & 1.34E-73 & 1.58E-251 & 8.14E-237 \\
$\uplambda_{\text{Edd}}$ & $\rho$ & 0.18 & 1 & 0.21 & -0.43 & 0.37 & 0.22 & 0.29 & -0.05 & 0.245 & -0.11 & 0.39 & 0.22 & -0.07 & -0.31 & -0.24 & 0.076 & 0.36 & 0.28 \\
 & P & 3.63E-15 & 0 & 2.93E-21 & 2.8E-84 & 3.73E-53 & 6.95E-07 & 4.1E-11 & 0.215 & 7.24E-08 & 0.012 & 4.41E-19 & 7.93E-07 & 0.004 & 3.8E-32 & 1.31E-21 & 0.003 & 9.46E-47 & 6.35E-28 \\
z & $\rho$ & 0.71 & 0.21 & 1 & 0.4 & 0.62 & 0.55 & 0.11 & 0.21 & 0.13 & -0.1 & 0.51 & 0.64 & X & -0.11 & 0.43 & 0.14 & 0.67 & 0.64 \\
 & P & 6.003E-286 & 2.93E-21 & 0 & 4.5E-73 & 2.63E-169 & 2.83E-40 & 0.012 & 2.49E-05 & 0.008 & 0.022 & 1.18E-32 & 6.42E-57 & X & 1.15E-05 & 2.04E-67 & 3.67E-08 & 4.55E-195 & 4.7E-171 \\
M$_{\text{\text{BH}}}$ & $\rho$ & 0.49 & -0.43 & 0.4 & 1 & 0.54 & 0.46 & -0.17 & 0.12 & 0.13 & 0.19 & 0.25 & 0.49 & -0.22 & -0.13 & 0.58 & X & 0.57 & 0.53 \\
 & P & 1.61E-112 & 2.8E-84 & 4.5E-73 & 0 & 2.11E-119 & 3.44E-26 & 0.0002 & 0.006 & 0.004 & 4.84E-05 & 2.08E-08 & 1.59E-29 & 1.12E-18 & 1.82E-07 & 4.94E-130 & X & 1.22E-82 & 5.3E-109 \\
L$_{3000\text{\AA}}$ & $\rho$ & 0.71 & 0.37 & 0.62 & 0.54 & 1 & 0.83 & X & X & 0.31 & X & 0.78 & 0.85 & -0.25 & -0.36 & 0.4 & 0.15 & 0.88 & 0.87 \\
 & P & 2.7E-217 & 3.73E-53 & 2.63E-169 & 2.11E-119 & 0 & 8.6E-110 & X & X & 2.7E-10 & X & 3.53E-78 & 1.99E-107 & 3.34E-23 & 8.68E-47 & 8.6E-57 & 1.8E-09 & 0 & 0 \\
L$_{5000\text{\AA}}$ & $\rho$ & 0.63 & 0.22 & 0.55 & 0.46 & 0.83 & 1 & X & -0.15 & 0.36 & 0.103 & 0.69 & 0.8 & -0.18 & -0.16 & 0.113 & 0.14 & 0.68 & 0.73 \\
 & P & 3.21E-52 & 6.95E-07 & 2.83E-40 & 3.44E-26 & 8.6E-110 & 0 & X & 0.002 & 7.7E-16 & 0.02 & 1.25E-66 & 1.28E-108 & 0.0004 & 0.001 & 0.02 & 0.004 & 1.44E-52 & 1.48E-63 \\
EW H$\beta_{\text{ILR}}$ & $\rho$ & X & 0.29 & 0.11 & -0.17 & X & X & 1 & 0.23 & 0.37 & X & 0.35 & X & 0.21 & X & -0.21 & X & X & X \\
 & P & X & 4.1E-11 & 0.012 & 0.0002 & X & X & 0 & 6.23E-07 & 7.48E-17 & X & 1.75E-14 & X & 2.46E-05 & X & 8.72E-05 & X & X & X \\
EW H$\beta_{\text{VBLR}}$ & $\rho$ & 0.21 & -0.05 & 0.21 & 0.12 & X & -0.15 & 0.23 & 1 & X & 0.12 & X & 0.35 & 0.16 & 0.23 & X & X & X & X \\
 & P & 2.71E-06 & 0.215 & 2.49E-05 & 0.006 & X & 0.002 & 6.23E-07 & 0 & X & 0.015 & X & 6.7E-15 & 0.002 & 1.83E-05 & X & X & X & X \\
FWHM H$\beta_{\text{ILR}}$ & $\rho$ & X & 0.245 & 0.13 & 0.13 & 0.31 & 0.36 & 0.37 & X & 1 & 0.48 & 0.54 & 0.34 & X & X & X & 0.14 & 0.29 & 0.26 \\
 & P & X & 7.24E-08 & 0.008 & 0.004 & 2.7E-10 & 7.7E-16 & 7.48E-17 & X & 0 & 1.24E-09 & 1.38E-37 & 4.18E-14 & X & X & X & 0.004 & 1.33E-08 & 8.14E-07 \\
FWHM H$\beta_{\text{VBLR}}$ & $\rho$ & X & -0.11 & -0.1 & 0.19 & X & 0.103 & X & 0.12 & 0.48 & 1 & X & 0.16 & -0.14 & 0.127 & 0.326 & X & X & 0.12 \\
 & P & X & 0.012 & 0.022 & 4.84E-05 & X & 0.02 & X & 0.015 & 1.24E-09 & 0 & X & 0.0003 & 0.004 & 0.014 & 0.041 & X & X & 0.03 \\
LH$\beta_{\text{ILR}}$ & $\rho$ & 0.63 & 0.39 & 0.51 & 0.25 & 0.78 & 0.69 & 0.35 & X & 0.54 & X & 1 & 0.72 & X & -0.25 & X & 0.17 & 0.74 & 0.64 \\
 & P & 1.53E-53 & 4.41E-19 & 1.18E-32 & 2.08E-08 & 3.53E-78 & 1.25E-66 & 1.75E-14 & X & 1.38E-37 & X & 0 & 5.35E-77 & X & 1.26E-06 & X & 0.0005 & 7.66E-68 & 6.41E-45 \\
LH$\beta_{\text{VBLR}}$ & $\rho$ & 0.71 & 0.22 & 0.64 & 0.49 & 0.85 & 0.8 & X & 0.35 & 0.34 & 0.16 & 0.72 & 1 & -0.1 & -0.106 & 0.154 & X & 0.74 & 0.79 \\
 & P & 6.23E-76 & 7.93E-07 & 6.42E-57 & 1.59E-29 & 1.99E-107 & 1.28E-108 & X & 6.7E-15 & 4.18E-14 & 0.0003 & 5.35E-77 & 0 & 0.04 & 0.04 & 0.002 & X & 5.72E-68 & 5.6E-82 \\
EW MgII$_{\text{core}}$ & $\rho$ & X & -0.07 & X & -0.22 & -0.25 & -0.18 & 0.21 & 0.16 & X & -0.14 & X & -0.1 & 1 & 0.41 & 0.17 & 0.26 & 0.16 & -0.06 \\
 & P & X & 0.004 & X & 1.12E-18 & 3.34E-23 & 0.0004 & 2.46E-05 & 0.002 & X & 0.004 & X & 0.04 & 0 & 9.05E-55 & 2.3E-10 & 3.08E-24 & 4.4E-09 & 0.009 \\
EW MgII$_{\text{wings}}$ & $\rho$ & X & -0.31 & -0.11 & -0.13 & -0.36 & -0.16 & X & 0.23 & X & 0.127 & -0.25 & -0.106 & 0.41 & 1 & X & 0.15 & -0.22 & 0.055 \\
 & P & X & 3.8E-32 & 1.15E-05 & 1.82E-07 & 8.68E-47 & 0.001 & X & 1.83E-05 & X & 0.014 & 1.26E-06 & 0.04 & 9.05E-55 & 0 & X & 3.18E-09 & 6.5E-18 & 0.032 \\
 FWHM MgII$_{\text{core}}$ & $\rho$ & X & -0.24 & 0.43 & 0.58 & 0.4 & 0.113 & -0.21 & X & X & 0.326 & X & 0.154 & 0.17 & X & 1 & 0.159 & 0.47 & 0.42 \\
 & P & X & 1.31E-21 & 2.04E-67 & 4.94E-130 & 8.6E-57 & 0.02 & 8.72E-05 & X & X & 0.041 & X & 0.002 & 2.3E-10 & X & 0 & 8.12E-10 & 7.5E-85 & 7.12E-67 \\
FWHM MgII$_{\text{wings}}$ & $\rho$ & 0.45 & 0.076 & 0.14 & X & 0.15 & 0.14 & X & X & 0.14 & X & 0.17 & X & 0.26 & 0.15 & 0.159 & 1 & 0.26 & 0.23 \\
 & P & 1.34E-73 & 0.003 & 3.67E-08 & X & 1.8E-09 & 0.004 & X & X & 0.004 & X & 0.0005 & X & 3.08E-24 & 3.18E-09 & 8.12E-10 & 0 & 3.5E-21 & 5.04E-20 \\
L MgII$_{\text{core}}$ & $\rho$ & 0.73 & 0.36 & 0.67 & 0.57 & 0.88 & 0.68 & X & X & 0.29 & X & 0.74 & 0.74 & 0.16 & -0.22 & 0.47 & 0.26 & 1 & 0.85 \\
 & P & 1.58E-251 & 9.46E-47 & 4.55E-195 & 1.22E-82 & 0 & 1.44E-52 & X & X & 1.33E-08 & X & 7.66E-68 & 5.72E-68 & 4.4E-09 & 6.5E-18 & 7.5E-85 & 3.5E-21 & 0 & 0 \\
L MgII$_{\text{wings}}$ & $\rho$ & 0.72 & 0.28 & 0.64 & 0.53 & 0.87 & 0.73 & X & X & 0.26 & 0.12 & 0.64 & 0.79 & -0.06 & 0.055 & 0.42 & 0.23 & 0.85 & 1 \\
 & P & 8.14E-237 & 6.35E-28 & 4.7E-171 & 5.3E-109 & 0 & 1.48E-63 & X & X & 8.14E-07 & 0.03 & 6.41E-45 & 5.6E-82 & 0.009 & 0.032 & 7.12E-67 & 5.04E-20 & 0 & 0 \\

\hline
\end{tabular}
\end{table}
\end{landscape}

\section*{Acknowledgements}
We acknowledge the Referee for providing useful comments which improved the quality of the work.
D. Nour acknowledges Al Baath University, Syria for the financial support.
KS acknowledges the financial support from the Core Research Grant programme under SERB, Government of India.\\
Funding for SDSS-III has been provided by the Alfred P. Sloan Foundation, the Participating Institutions, the National Science Foundation, and the U.S. Department of Energy Office of Science. The SDSS-III web site is http://www.sdss3.org/.

SDSS-III is managed by the Astrophysical Research Consortium for the Participating Institutions of the SDSS-III Collaboration including the University of Arizona, the Brazilian Participation Group, Brookhaven National Laboratory, Carnegie Mellon University, University of Florida, the French Participation Group, the German Participation Group, Harvard University, the Instituto de Astrofisica de Canarias, the Michigan State/Notre Dame/JINA Participation Group, Johns Hopkins University, Lawrence Berkeley National Laboratory, Max Planck Institute for Astrophysics, Max Planck Institute for Extraterrestrial Physics, New Mexico State University, New York University, Ohio State University, Pennsylvania State University, University of Portsmouth, Princeton University, the Spanish Participation Group, University of Tokyo, University of Utah, Vanderbilt University, University of Virginia, University of Washington, and Yale University.

\section*{Data Availability}
Data used in this work can be accessed from https://www.sdss.org/dr14/ and is also available with authors



\bibliographystyle{mnras}
\bibliography{references} 





\bsp	
\label{lastpage}
\end{document}